\documentclass[pre,aps,epsf,amsmath,amsfonts]{revtex4}

\usepackage{stmaryrd}
\usepackage{graphicx}

\newcommand{\beq}{\begin{equation}}
\newcommand{\eeq}{\end{equation}}
\begin{document}
\title{Dynamics and geometric properties of the k-Trigonometric model}
\author{
  F.~Zamponi$^{1}$, L.~Angelani$^{1,2}$, L.~F.~Cugliandolo$^{3,4}$, J.~Kurchan$^{5}$, G.~Ruocco$^{1}$
        }
\affiliation{
  $^1$Dipartimento di Fisica and INFM, Universit\`a di Roma {\em La Sapienza},
  P. A. Moro 2, 00185 Roma, Italy \\
  $^2$SMC - INFM 
  Universit\`a di Roma {\em La Sapienza}, P. A. Moro 2, 00185 Roma, Italy\\
  $^3$ Laboratoire de Physique Th\'eorique, Ecole Normale Sup\'erieure, 24 rue Lhomond
  75231 Paris Cedex 05 France \\
  $^4$ Laboratoire de Physique Th\'eorique et Hautes Energies
  Jussieu,
 4, Place Jussieu,  75252 Paris Cedex 05 France \\
$^5$ P.M.M.H. Ecole sup\'erieure de Physique et Chimie industrielles, 10, Rue Vauquelin 75231 Paris
 Cedex 05, France
}
\date{\today}
\begin{abstract}
We analyze the dynamics and the geometric properties of the 
Potential Energy Surfaces (PES)
of  the $k$-Trigonometric  Model
($k$TM), defined by a fully-connected $k$-body
interaction.
This model has no thermodynamic transition for $k=1$, a
second order one for $k$=2, and a first order one for $k$$>$2.
In this paper we
{\it i)}~show that  the single particle dynamics can be traced back to an effective 
dynamical system (with only one degree of freedom);
{\it ii)}~compute the diffusion constant analytically;
{\it iii)}~determine analytically several properties of the self correlation functions 
apart from the relaxation times which we calculate numerically;
{\it iv)}~relate the collective correlation functions
to the ones of the effective degree of freedom using an 
exact Dyson-like equation;
{\it v)}~using two analytical methods, calculate 
the saddles of the PES that are visited by the system evolving at fixed 
temperature. On the one hand we minimize $|\nabla V|^2$, as usually done in the numerical study
of supercooled liquids and, on the other hand, we compute the
saddles with minimum distance (in configuration space) from initial equilibrium 
configurations. We find the same result from the two calculations and
we speculate that the coincidence might go beyond the specific model investigated here.
\end{abstract}
%\pacs{05.20.-y, 31.50.-x, 75.10.Hk}
\maketitle

%%%%%%%%%%%%%%%  TEXT  %%%%%%%%%%%%%%%%

\section{Introduction}

\noindent
In the last few years there has been an intensive study 
\cite{deb_nature,land_angell,land_sastry,land_buc,land_sastry2,land_keyes,sorin}
concerning the connection between the slow
dynamics of complex systems and the extrema of the potential energy surface (PES),
defined through the potential energy function $V({q})$,
following the seminal 
work of Stillinger and Weber~\cite{stillinger}.

Two ways of studying the dynamics of supercooled liquids and glasses that are based on the 
analysis of the PES
have been of particular importance in the recent past. The first one concerns 
the analysis of the properties (energy location, number, curvature,..) of the minima of 
the PES that are visited by the system during its evolution at fixed
temperature. Assigning to each minimum its zero-temperature basin of attraction, one obtains
a partition of the phase space allowing to define a configurational entropy for the
supercooled and the out-of-equilibrium glassy regime~\cite{fs_entropy}.
The properties of the minima of the PES 
have also been connected to several features of supercooled liquids and glasses. 
We can mention their relation  to the {\it fragility} of the glass former~\cite{land_sastry2}, 
the diffusion processes in supercooled liquids~\cite{land_keyes,fabr,la_98,donati}, 
and the effective fluctuation-dissipation 
temperature~\cite{FDTgen,Cukupe,cugliandolo2} in the out-of-equilibrium glassy phase~\cite{fs_aging}.
This method is closely related to Edwards' proposal~\cite{Edwards} to describe the main properties of 
granular matter  with a {\it flat} measure over blocked configurations that correspond to 
the minima of the PES. Note that granular matter is an effectively zero-temperature system
for which the study of the energy landscape if fully justified.  

The second approach, which corresponds to the study of the {\it instantaneous normal modes}, 
is based on the study of the eigenvalues of the Hessian at the instantaneous 
configurations along the trajectory (in configurational space) 
that the system follows during its dynamical evolution~\cite{keyes_inm,keyes_vari}.
This approach allowed one to relate the diffusion process to the local curvature of the landscape. 

More recently a third approach has been proposed~\cite{noi_sad,cav_sad} and 
applied~\cite{sad_1,doye,sad_3,grig,sadBLJ,parisi_boson} to study the slow dynamics in 
supercooled liquids. Within this approach, the {\it saddles} of the potential energy 
surface play a central role. It has been found numerically that the order of the saddles (number of 
negative eigenvalues of the Hessian matrix) visited during the equilibrium 
dynamics at temperature $T$ extrapolates to zero when $T$ reaches the dynamic transition 
temperature $T_{MCT}$ (or mode-coupling temperature~\cite{mct}).

The role of the stationary points of the PES (saddles and minima) has
been also 
pointed out in a different context. Indeed, studies aiming to clarifying the 
microscopic origin of phase transitions suggest that the presence and order of such
transitions is related to changes in the topology of the manifold of the 
PES sampled by the system when crossing the (thermodynamic) critical 
point~\cite{pettini&co,noieulero}. This has been observed by counting the number and the order 
of the stationary points of $V(q)$ and building up the Euler characteristic of the manifold. 
The latter is a genuine topological property of the energy surface defined by $V(q)=$constant, 
and, in particular, it does not depend on the statistical measure defined on it (i.e., on temperature).

Disordered mean-field spin models have been proposed to mimic the behaviour of super-cooled liquids and 
glasses. Their statics and dynamics, as well as the properties of their free-energy and 
energy landscapes, are amenable to analytical studies~\cite{cugliandolo2}. The main features mentioned in the
previous paragraphs are realized by these models where, at finite temperature, the geometry of the 
free-energy landscape replaces the PES. In particular, the importance of saddles in the 
free-energy landscape for the evolution of these systems has been elucidated in the past 
and a comparison between the roles played by free-energy and energy landscapes has also 
been discussed in this and more general contexts~\cite{Kula,Bimo}.

If one wishes to relate the equilibrium dynamics of a complex system to the properties of the 
saddles of its PES, an unambiguous definition of {\it saddle visited during the equilibrium dynamics}
is mandatory. Until now, two different definitions have been used:
{\it 1)} In the numerical simulations of simple models -- but still too hard to study analytically --  
such as Lennard-Jones systems, a partitioning 
of the configuration space in basins of attraction of saddles is obtained via an appropriate 
function $W$ (usually $W=|\nabla V|^2$) that has a local minimum on each stationary point of $V$, 
and the saddles are then obtained via a minimization of $W$ starting from an equilibrium configuration 
obtained from a molecular dynamics simulation at temperature $T$.
{\it 2)} In the analytic computations 
applied to disordered mean-field spin models
one looks to the saddles that are closest, with respect to some distance in the configuration space, 
to a reference configuration extracted from the Gibbs distribution at temperature $T$~\cite{cav2}. 
Unfortunately, until now
the two methods have been applied to different models so the comparison between them is still 
qualitative.

In this paper we study a very simple mean-field model without
disorder, the $k$ Trigonometric Model ($k$TM), for which one can
calculate analytically all the 
relevant quantities that have been previously 
studied numerically for more realistic models. In spite of its simplicity, the thermodynamic 
behavior of this model is quite rich, and its PES shows some of the features that have been 
observed in the PES of Lennard-Jones systems \cite{noi_sad,cav_sad,sad_1,doye,sad_3,grig,sadBLJ}.  
Unfortunately, the model is too simple to show an interesting dynamics. The dynamical behaviour 
is closely related to the thermodynamics and, due to the absence of frustration or disorder, no dynamical arrest
is observed. Still, on the one hand we have been able to
check analytically the validity of some ideas that had been proposed in the literature and, on the other 
hand, to elaborate a method for the minimization of $W=|\nabla V|^2$ that will be of use for a larger class of 
mean-field models including the disordered ones. This method allowed us to compare analytically the 
two definitions of closest saddles to equilibrium configurations, and to show that they coincide 
in our model. 

The paper is organized as follows: in section \ref{sec:model} we introduce the 
model and present its main features. Then we analyze its static properties: in 
section \ref{sec:thermo} we discuss its thermodynamical behavior, in section \ref{topologia} 
we study the topological properties of the PES and relate them to the results of 
section \ref{sec:thermo}. In sections \ref{sec:single_dyn} and \ref{sec:coll_dyn} we present 
a detailed study of the dynamical behavior of the model. Then, in section \ref{selleW} we 
discuss the definition of the closest saddles to equilibrium configurations, their properties 
and their relation with the dynamics of the system. Many of the calculations require the 
introduction of a formalism that may not be familiar to all the readers and is not really 
necessary to follow the relevant part of the presentation; they are then presented in detail
in the appendices.

%%%%%%%%%%%%%%%%%%%%%%%%%%%%%%%%%%%%%%%%%%%%%%%%%%%%%%%%%%%%%%%%%%%%%%%%%%%%%%%%%%%%%%%%%%%%%%%%%%%%%%%%%%%%%%%%%%%%%%%%%%%%%%%%%%%%%%%%%%%%%%%%%%%%%%%%%%%%%%%

\section{The model}
\label{sec:model}

\noindent
The $k$-Trigonometric Model ($k$TM) has been introduced in \cite{noieulero} with the aim to study the relation between phase transitions and topological properties of the PES. The model is defined by the Hamiltonian
\beq
\label{hamiltonian}
H=\frac{\Delta}{N^{k-1}} \sum_{i_1,\cdots,i_k} [1-\cos(\varphi_{i_1}+\cdots+\varphi_{i_k})] = N \Delta (1-\Re z^k) \ ,
\eeq
having introduced the ``magnetization''
\beq
\label{zetadef}
 z=\frac{1}{N} \sum_i e^{i\varphi_i}=\xi e^{i \psi} \ .  \eeq Here
$\varphi_i$$\in$$[0,2\pi)$, $i$$=$$1,\cdots,N$, are angular variables
and $\Delta$ is the energy scale.  It is easy to see that the model is
not invariant under continuous transformations of $\varphi_i$ but only
under the discrete group $C_{kv}$ generated by
\beq
\label{symm}
\begin{split}
&\varphi_i \rightarrow \varphi_i +\frac{2 \pi}{k} \ , \\
&\varphi_i \rightarrow -\varphi_i \ .
\end{split}
\eeq If one interprets the variable $\varphi_i$ as the angle between a
two-dimensional unitary vector and a fixed axis, the trasformations in
Eq.~(\ref{symm}) are rotations by an angle $2\pi/k$  of the vector and
the reflection with respect to the fixed axis. In the low temperature phase
this symmetry is broken, and a spontaneous magnetization is generated
in a direction $\psi_n = 2 \pi n/k$. We will often choose $\psi=0$
without loss of generality, in order to keep unbroken the symmetry
$\varphi_i \rightarrow -\varphi_i$ \cite{nota1}. \\ 
The system is
subject to a dynamics generated by a Langevin equation \cite{ZJ}: \beq
\label{dinamicalang}
\gamma \dot{\varphi_i}=-\frac{\partial H}{\partial \varphi_i} + \eta_i \ ,
\eeq
where $\gamma/\Delta$ is the time scale and $\eta_i$ is a Gaussian noise with
\beq
\begin{split}
&\langle \eta_i(t) \rangle = 0 \ , \\
&\langle \eta_i(t) \eta_j(t') \rangle = 2T \gamma \delta_{ij}\delta(t-t') \ .
\end{split}
\eeq
We will consider averages of a generic observable $A( \{ \varphi \})$
over the noise distribution, and we choose random initial data (that
corresponds to a quench from infinite temperature). Eventually, we
will consider the $t$$\rightarrow$$\infty$ limit, in which the system
equilibrates and is described by the Gibbs ensemble at temperature
$T$.

%%%%%%%%%%%%%%%%%%%%%%%%%%%%%%%%%%%%%%%%%%%%%%%%%%%%%%%%%%%%%%%%%%%%%%%%%%%%%%%%%%%%%%%%%%%%%%%%%%%%%%%%%%%%%%%%%%%%%%%%%%%%%%%%%%%%%%%%%%%%%%%%%%%%%%%%%%%%%%%%%%%%%%%%%%%%%%%%%%%%%%%%%%%%%%%%%%%%%%%%%%%%%%%

\section{Thermodynamics}
\label{sec:thermo}

\noindent
The termodynamics of mean field models is exactly solved neglecting
the correlations between different degrees of freedom and obtaining an
effective Hamiltonian that contains a parameter to be determined
self-consistently. For example, in the fully connected Ising model,
with Hamiltonian $H=-(2N)^{-1} \sum_{ij} s_i s_j$, the substitution
reads $s_i s_j \rightarrow \langle s_i \rangle s_j + s_i \langle s_j
\rangle - \langle s_i \rangle \langle s_j \rangle$. Defining
$m=\langle s_i \rangle$, one obtains the effective Hamiltonian ${\cal
H}(s) = m s + c(m)$ for a single degree of freedom $s$ ($c(m)$ is an
irrelevant constant that depends only on $m$). The self-consistency
equation is finally obtained calculating $m = \langle s \rangle$ on
this effective Hamiltonian. In fact, one can show that this procedure
is equivalent to the evaluation of the free energy at the saddle point
in the $N \rightarrow \infty$ limit. \\ 
%%%%%%%%%%%%  THERMO 1 %%%%%%%%%%%%%%%%%%%%%%%%
\begin{figure}[t]
\centering
\includegraphics[width=.47\textwidth,angle=0]{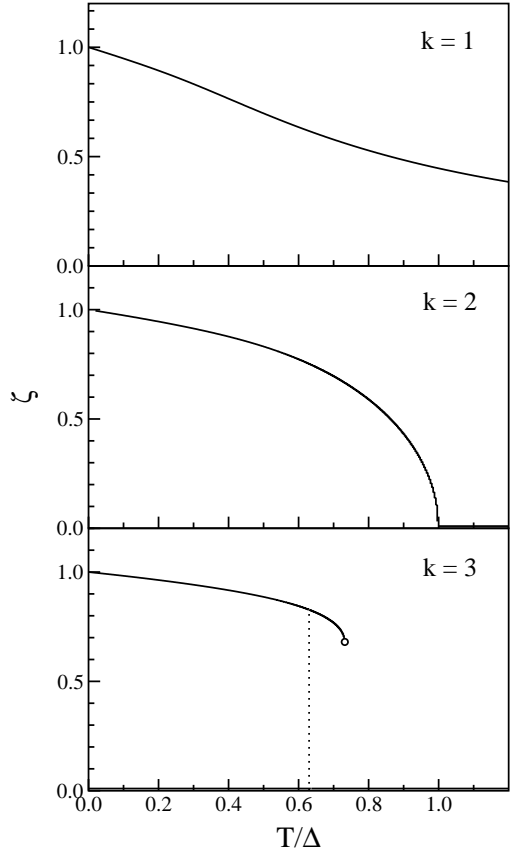}
\caption{
Mean magnetization $\zeta$ as a function of temperature for $k\!=\!1,2,3$.
For $k\!=\!3$ the value of the canonical transition temperature $T_0\!=\!0.63$ is indicated by a dotted vertical line, while the temperature $T_c=0.72$ at which the magnetic solution disappears is marked by a white dot.
The magnetic solution is metastable for $T_0<T<T_c$; the same happens to the $\zeta=0$ solution for $T<T_0$.}
\label{thermo1}
\end{figure}
%%%%%%%%%%%%%%%%%%%%%%%%%%%%%%%%%%%%%%%%%%%%%%
The generalization of this
procedure to the $k$TM is obtained substituting in
Eq.~(\ref{hamiltonian}) the expression: \beq e^{i\varphi_{i_1}} \cdot
\cdot \cdot e^{i\varphi_{i_k}} \rightarrow k \ e^{i\varphi_{i_1}}
\langle e^{i\varphi_{i_2}}\rangle \cdot \cdot \cdot \langle
e^{i\varphi_{i_k}} \rangle - (k-1) \ \langle e^{i\varphi_{i_1}}\rangle
\cdot \cdot \cdot \langle e^{i\varphi_{i_k}} \rangle \ .  \eeq and
introducing the mean (complex) ``magnetization'' $\zeta\!=\!\langle
e^{i\varphi}\rangle$, that has to be determined self-consistently on
the mean field effective Hamiltonian ${\cal H}$. As we always chose
$\psi=0$, $\zeta$ is real and the effective Hamiltonian reads:
\begin{equation}
{\cal H} = \Delta [ 1 + (k-1) \zeta^{k} - k \zeta^{k-1} \cos \varphi ] \ .
\label{Heff}
\end{equation}
The self consistency equation for $\zeta$ turns out to be:
\begin{equation} 
\zeta  = \langle \cos \varphi \rangle _{\cal H} =
\frac{I_1 (\beta \Delta k \zeta^{k-1})}{I_0 (\beta \Delta k \zeta^{k-1})} \ ,
\label{zeta}
\end{equation}
where $\beta\!=\!1/T$, $I_0 (\alpha) = (2\pi)^{-1}\int_0^{2\pi}
d\varphi e^{\alpha \cos \varphi}$ and $I_1 (\alpha) = I_0^{'}
(\alpha)$ are the modified Bessel functions of order 0 and 1
respectively.  For each $\beta$ the Eq.~(\ref{zeta}) gives the
thermodynamic value of the mean magnetization $\zeta (T)$.  The value
$\zeta\!=\!0$ always solves Eq.~(\ref{zeta}), but is a stable solution
only at low $\beta$ (high temperature).
%%%%%%%%%%%%   THERMO 2   %%%%%%%%%%%%
\begin{figure}[t]
\centering
%\vspace{1cm}
\includegraphics[width=.47\textwidth,angle=0]{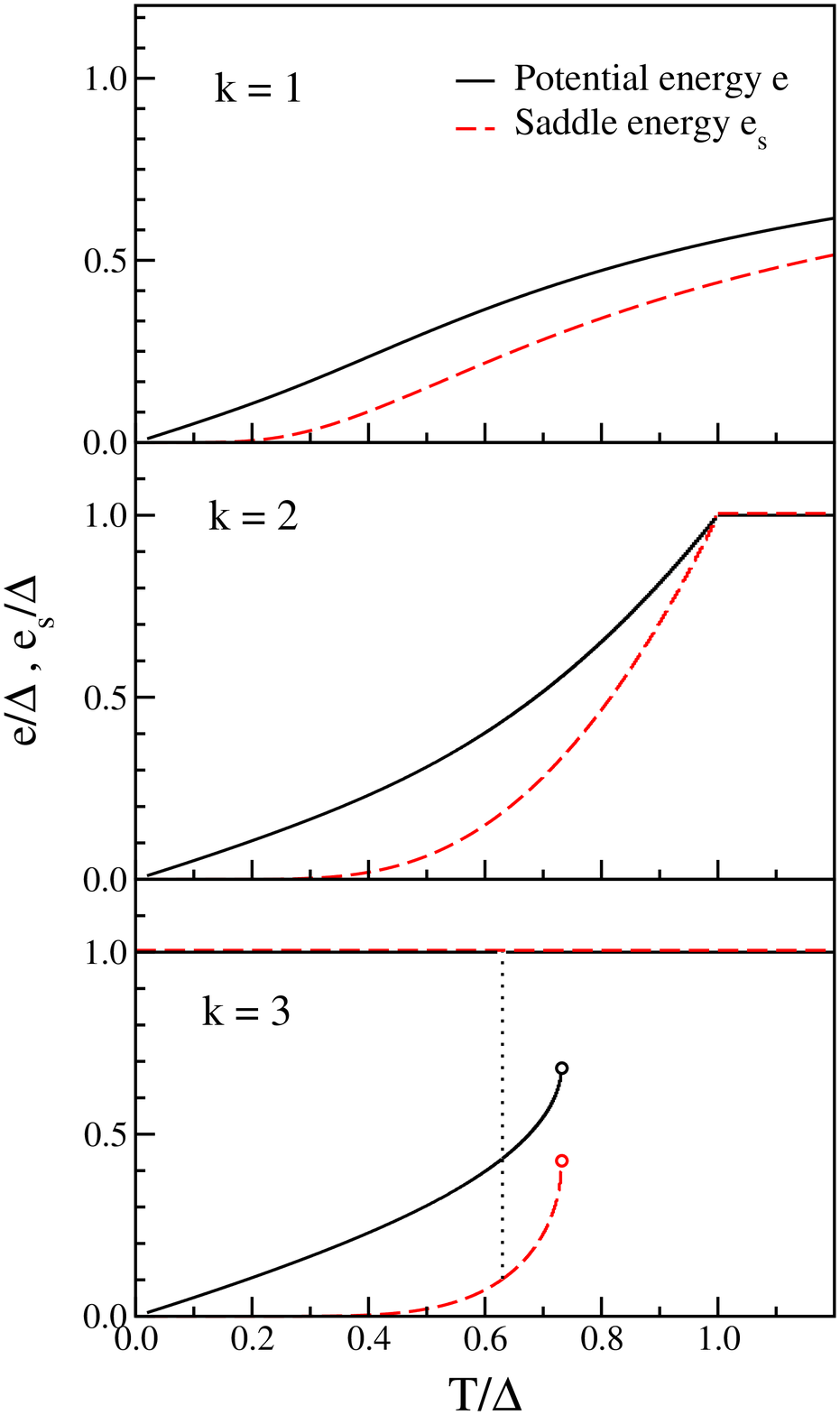}
\caption{Potential energy $e$ (full lines) and saddle energy $e_s$ (dashed lines, see section \ref{selleW})
 as a function of temperature
for $k\!=\!1,2,3$. The notation is the same as in the previous figure.
}
\label{thermo2}
\end{figure}
%%%%%%%%%%%%%%%%%%%%%%%%%%%%%%%%%%%%%%
 As $\beta$ is increased other
solutions may appear, and one has to consider the one that minimizes
the free energy as the stable one, while the other solutions can be
either unstable or metastable.  In Fig.~\ref{thermo1} we report the
function $\zeta(T)$ for $k\!=\!1,2,3$.  For $k\!=\!1$ the curve is
smooth, no phase transition occurs and the magnetization approaches
zero at high temperature.  For $k\!=\!2$ a second order phase
transition takes place at $T_c=\Delta$, separating a high temperature
paramagnetic phase, where only the solution $\zeta\!=\!0$ exists, and
a low temperature ordered phase, where $\zeta=0$ becomes a maximum of
the free energy separating two minima with $\zeta \neq 0$
corresponding to $\psi\!=\!0$ and $\psi\!=\!\pi$.  From this symmetry
structure one sees that for $k=2$ the model is in the universality
class of the Ising model; the (scalar) order parameter is the real
part of the magnetization, while the imaginary part is never different
from zero. The critical exponents are then the classical mean-field
exponents of the Ising model; in particular we have $\zeta \sim |T -
T_c|^{1/2}$ close to $T_c$ and, if a perturbation $\delta {\cal H} = -
h \cos \varphi $ is added, $\zeta \sim h^{1/3}$ at $T=T_c$.  For
$k\!=\!3$ (and also for $k\!>\!3$, not reported in the figure) the
system undergoes a first order phase transition. At high $T$ only the
paramagnetic solution $\zeta\!=\!0$ exists, but on lowering $T$ two
other solutions appear at $T=T_c$: the magnetic one with $\zeta>0$ and
degeneration $k$ (i.e. $k$ different possible values of $\psi$) and
the one corresponding to the maximum of the free energy separating the
magnetic and paramagnetic solutions. The magnetic minimum becomes
stable at $T=T_0<T_c$.  In Fig.~\ref{thermo1}, for the $k\!=\!3$ case,
we report the evolution of the two minima as a function of
temperature.  The second minimum appears at $T_c = 0.72$ (white dot in
the figure) and the transition temperature $T_0 = 0.63$ is indicated
by a dotted vertical line.  In Fig.~\ref{thermo2} the temperature
dependence of the potential energy $e=\langle {\cal H} \rangle =
\Delta (1-\zeta^k)$ (full lines in the figures) is shown for
$k\!=\!1,2,3$. The first order phase transition also manifests itself
in a discontinuity of the potential energy as a function of
temperature.

%%%%%%%%%%%%%%%%%%%%%%%%%%%%%%%%%%%%%%%%%%%%%%%%%%%%%%%%%%%%%%%%%%%%%%%%%%%%%%%%%%%%%%%%%%%%%%%%%%%%%%%%%%%%%%%%%%%%%%%%%%%%%%%%%%%%%%%%%%%%%%%%%%%%%%%%%%%%%%%

\section{Topological properties of the energy surface}
\label{topologia}

\noindent
In this section we will study the properties of the stationary points
(saddles) of the Potential Energy Surface (PES) of the system, defined by
the Hamiltonian (\ref{hamiltonian}). We will now focus only on the
{\it topological} properties of the saddles, while in section
\ref{selleW} we will study the properties of the saddles sampled by
the system equilibrated at temperature $T$. We will now follow the
derivation in \cite{noieulero}, while in appendix \ref{app_euler} we
present a different derivation that will be useful in the following.
The stationary points $\bar{\varphi}$ are defined by the condition
$dH(\bar{\varphi})$$=$$0$, and their order $\nu$ is defined as the
number of negative eigenvalues of the Hessian matrix
$\text{H}_{ij}(\bar{\varphi})$$=$$({\partial^2 H}/{\partial \varphi_i
\partial \varphi_j})|_{\bar{\varphi}}$.  To determine the location of
the stationary points we have to solve the system 
\beq \frac{\partial
H}{\partial \varphi_j}=-\Delta \ k \ \Re[i z^{k-1} e^{i \varphi_j}] =
\Delta \ k \ \xi^{k-1} \sin[(k-1)\psi + \varphi_j] = 0 \ , \ \forall j
\ ,
\label{saddef}
\eeq where we have used the definition $z$$\equiv$$\xi e^{i\psi}$
given in Eq.~(\ref{zetadef}).  A first group of solutions arises for
$\xi=0$; from Eq.~(\ref{hamiltonian}) we have
$H(\varphi)$$=$$N\Delta[1-\xi^k \cos(k\psi)]$, and then the stationary
points with $\xi(\bar{\varphi})$=0 are all located at the energy
$e$=$H(\bar{\varphi})/N$=$\Delta$. We will now restrict ourselves to
the region $e\neq \Delta$ because, as we will see at the end, the
quantities in which we are interested are singular when $e=\Delta$. The
presence of this singularity seems to be related, as we discussed
elsewhere \cite{noieulero}, to the presence (and the order) of a phase
transition.  For $e \neq \Delta$, Eq.~(\ref{saddef}) becomes
\begin{equation}
\sin [(k-1)\psi + \varphi_j]=0 \ , \ \forall j \ ,
\end{equation}
and its solutions are
\begin{equation}
\label{sadcond}
\bar{\varphi}_j^{\mathbf{m}}=[m_j \pi - (k-1) \psi]_{\text{mod} \ 2 \pi} \ ,
\end{equation}
where $m_j$$\in$$\{ 0,1 \}$ and $\mathbf{m}\equiv\{m_j \}$. Therefore, beside the different possible values of $\psi$, each stationary point
$\bar{\varphi}^{\mathbf{m}}$ is characterized by the set
$\mathbf{m}$. To determine the unknown constant
$\psi$ we have to substitute Eq.~(\ref{sadcond}) in the
self-consistency equation
\begin{equation} \label{selfcons}
z = \xi e^{i\psi} = N^{-1}
\text{$\sum$}_j
e^{i\varphi_j}=N^{-1} e^{-i\psi(k-1)}
\text{$\sum$}_j (-1)^{m_j} \ .
\end{equation}
If we introduce the quantity $n(\bar{\varphi})$ defined by
\begin{equation}
\label{frac_ord}
n=N^{-1}
\text{$\sum$}_j
 m_j \ ,
\hspace{1cm}
 1-2n=N^{-1}
\text{$\sum$}_j
 (-1)^{m_j} \ ,
\end{equation}
and we have from Eq.~(\ref{selfcons})
\begin{eqnarray}
&\label{z-x}
\xi=|1-2n| \ , \\
&\psi_l=
\begin{cases}
2l\pi /k \hspace{2cm} \text{ for } n < 1/2 \ , \\
(2l+1)\pi /k \hspace{1.12cm} \text{ for } n > 1/2 \ ,
\end{cases}
\label{rhoepsi}
\end{eqnarray}
where $l \in$$\mathbb{Z}$. Then the choice of the set $\mathbf{m}$
is not sufficient to specify the set $\{\varphi_j\}$ because the
constant $\psi$ can assume some different values. This fact is
connected with the symmetry structure of the potential energy
surface (the different values of $\psi_l$ generate the multiplets
of stationary points). We have then obtained that all the stationary points
 of energy
$e$$\neq$$\Delta$ have the form
\begin{equation}
\label{sadfin}
\bar{\varphi}_j^{\mathbf{m},l}=[m_j \pi - (k-1)
\psi_l]_{\text{mod} \ 2 \pi} \ .
\end{equation}
The Hessian matrix is given by
\beq
\text{H}_{ij}=
\Delta\ k \ \Re [N^{-1} (k-1) z^{k-2} e^{i(\varphi_i+\varphi_j)} 
+\delta_{ij} z^{k-1} e^{i \varphi_i)} ]  \ .
\eeq
%%%%%%%%%%  SIGMA  %%%%%%%%%%%%
\begin{figure}[t]
\centering
%\vspace{1cm}
\includegraphics[width=.47\textwidth,angle=0]{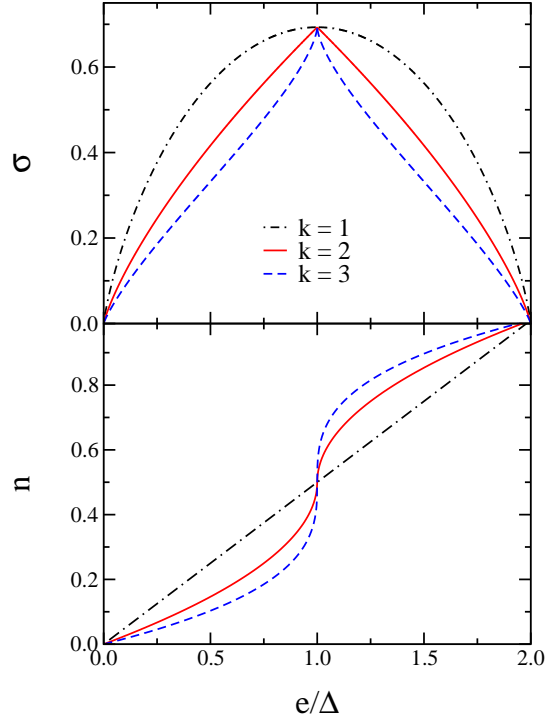}
\caption{The logarithm $\sigma$ of the number of saddles and the saddle order $n$ as a function of the energy level $e$ for $k\!=\!1,2,3$.}
\label{sigma}
\end{figure}
%%%%%%%%%%%%%%%%%%%%%%%%%%%%%%%%%%
In the thermodynamic limit it becomes diagonal
\begin{equation}
\label{hessdiag}
\text{H}_{ij}=
\delta_{ij}\ \Delta\ k\ \xi^{k-1}\
\cos \left( \psi(k-1)+\varphi_i \right) \ .
\end{equation}
One cannot {\it a priori} neglect the contribution of the
off-diagonal terms to the eigenvalues of $\text{H}$, but we have
numerically checked that their contribution changes the sign of at most
 one eigenvalue out of $N$. Neglecting the off-diagonal
contributions, the eigenvalues $\lambda_j$ of the Hessian calculated at the
stationary point $\bar{\varphi}$ are obtained substituting Eq.~(\ref{sadfin})
in Eq.~(\ref{hessdiag}):
\begin{equation}
\lambda_j = (-1)^{m_j} \Delta\ k\ \xi^{k-1} \ ,
\end{equation}
so the stationary point order $\nu(\bar{\varphi})$, defined as the number of negative eigenvalues of the Hessian matrix, is simply the number of $m_j$=1 in the set
$\mathbf{m}$ associated with $\bar{\varphi}$; we can identify the quantity $n(\bar{\varphi})$ given
by Eq.~(\ref{frac_ord}) with the fractional order $\nu(\bar{\varphi})/N$ of
$\bar{\varphi}$. Then, from Eqs.~(\ref{hamiltonian}), (\ref{z-x}) and
(\ref{rhoepsi}) we get a relation between the fractional order
$n(\bar{\varphi})$ and the potential energy
$e(\bar{\varphi})$$=$$H(\bar{\varphi})/N$ at each stationary  
point $\bar{\varphi}$. It reads:
\begin{equation}
\label{x-v}
n(e)=\frac{1}{2}
\left[1-\text{sgn}\left(1-\frac{e}{\Delta} \right)
\left|1-\frac{e}{\Delta}\right|^{1/k} \right] \ ,
\end{equation}
Moreover, the number of stationary points of given order $\nu$ is simply the
number of ways in which one can choose $\nu$ times 1 among the
$\{m_j \}$, see Eq.~(\ref{sadfin}), multiplied by a constant
$C_k$ that does not depend on $N$ and takes into account the degeneracy
introduced by
Eq.~(\ref{rhoepsi}). Therefore: $i)$ the fractional order
$n$$=$$\nu/N$ of the stationary points is a well defined monotonic function
of their potential energy $e$, given by Eq.~(\ref{x-v}), and
$ii)$ the number of stationary points of a given order $\nu$ is $C_k 
{N \choose \nu}$. We can define the quantity
\beq
\sigma(e) = \lim_{N \rightarrow \infty} \frac{1}{N} \log {N \choose Nn(e)} =
 -n(e) \log n(e) - (1-n(e)) \log (1-n(e)) \ ,
\eeq
that represents the ``configurational entropy'' of the saddles. In \cite{noieulero} we have shown that this quantity is related to the Euler characteristic of the manifolds $M_e=\{\varphi | H(\varphi) \leq Ne \}$ and that its singular behavior around the point $e=\Delta$ is related to both the presence and the order of the phase transitions that occur for $k \geq 2$. In Fig. \ref{sigma} the quantity $\sigma(e)$ and $n(e)$ are reported for $k=1,2,3$ for all values of $e \neq \Delta$: one can see that the presence of a phase transition for $k$$\geq$$2$ is signaled by a singularity in the first derivative of $\sigma(e)$. The order of the transition seems to be related to the sign of the second derivative of $\sigma(e)$ around the transition point, that is negative for second order phase transitions and positive for first order ones.

%%%%%%%%%%%%%%%%%%%%%%%%%%%%%%%%%%%%%%%%%%%%%%%%%%%%%%%%%%%%%%%%%%%%%%%%%%%%%%%%%%%%%%%%%%%%%%%%%%%%%%%%%%%%%%%%%%%%%%%%%%%%%%%%%%%%%%%%%%%%%%%%%%%%%%%%%%%%%

\section{Single-particle dynamics}
\label{sec:single_dyn}

\noindent
The single-particle dynamics can be studied by means of observables of the form
\beq
\label{onepartobs}
A ( \{ \varphi \} ) =\frac{1}{N} \sum_i {\cal A}(\varphi_i) \ .
\eeq
Some interesting quantities are, for example,
the diffusion constant $D(T)$, defined as
\beq
\label{defdiff}
D(T)=\lim_{t \rightarrow \infty} \frac{1}{2Nt} \sum_i \langle | \varphi_i(t)-\varphi_i(0) |^2 \rangle \ ,
\eeq
(where obviously the angular variables have to be considered as variables defined on the whole real axis without introducing the periodic condition $\varphi_i + 2\pi = \varphi_i$) and the self correlation function, defined as
\beq
\label{incoerente}
F(t,t')= \frac{1}{N} \sum_i \left[ \langle e^{i[\varphi_i(t) - \varphi_i(t')]} \rangle - \langle e^{i \varphi_i(t)} \rangle \langle e^{-i \varphi_i(t')} \rangle \right] \ . 
\eeq
To study the single-particle dynamics, we introduce an effective dynamical system for a single degree of freedom $\varphi$, defined by
\beq
\label{effdynsyst}
\begin{cases}
{\cal H}(\varphi,t) = - \Delta \Re [k \zeta(t)^{k-1} e^{i \varphi}] \ , \\
\gamma \dot{\varphi}(t) =-\frac{\partial {\cal H}}{\partial \varphi}(\varphi(t),t) + \eta(t) \ , \\
\zeta(t) = \langle e^{i \varphi(t)} \rangle \ ,
\end{cases}
\eeq
where again $\eta$ is a $\delta$-correlated Gaussian noise with
variance $2\gamma T$ and the averages are calculated on its
distribution. Note that the derivative of the effective Hamiltonian
${\cal H}(\varphi,t)$ is performed with respect to $\varphi$ at fixed
time $t$ (the time-dependence of ${\cal H}$ is encoded in
$\zeta(t)$). The last equation is, as in the static case, a
self-consistency equation.  In appendix \ref{app_self} we show that
the averages of observables of the type (\ref{onepartobs}) on the
dynamics defined by Eq.~(\ref{dinamicalang}) can be calculated using this
effective single-variable dynamical system: 
\beq
\label{legameeffettivo}
\langle A( \{ \varphi \} ) \rangle = \langle {\cal A}(\varphi)
\rangle_{\cal H} \ .  
\eeq 
We will consider the limit $t \rightarrow
\infty$ in which the system is in equilibrium, and $\zeta$ does not
depend on $t$ and is equal to its equilibrium value $\zeta(\beta)$
given by Eq.~(\ref{zeta}). In this limit the correlation function
(\ref{incoerente}) depends only on the time difference,
$F(t,t')=F(t-t')$, and the dynamical system (\ref{effdynsyst}) reduces
to \beq
\label{effdyneq}
\gamma \dot{\varphi}(t) + k \Delta \zeta(\beta)^{k-1} \sin \varphi(t) = \eta(t) \ .
\eeq
If $\zeta \neq 0$, we can define the reduced variables $\tilde{t} \equiv \frac{k\Delta\zeta^{k-1}}{\gamma} \ t$ and $\tilde{\eta}(\tilde{t}) \equiv \frac{1}{k \Delta \zeta^{k-1}} \eta(t)$, and Eq.~(\ref{effdyneq}) can be rewritten as
\beq
\label{effdynresc}
\frac{d\varphi}{d\tilde{t}} + \sin \varphi(\tilde{t}) = \tilde{\eta}(\tilde{t}) \ ,
\eeq
with $\langle \tilde{\eta}(\tilde{t}) \tilde{\eta}(0) \rangle = \frac{2T}{k \Delta \zeta^{k-1}} \delta(\tilde{t}) \equiv 2 \tilde{T} \delta(\tilde{t})$. The $k$ dependence is then encoded in $\tilde{t}$ and $\tilde{T}.$

%%%%%%%%%%%%%%%%%%%%%%%%%%%%%%%%%%%%%%%%%%%%%%%%%%%%%%%%%%%%%%%%%%%%%%%%%%%%%%%

\subsection{The diffusion constant}

%%%%%%%%%%  DIFFUSION  %%%%%%%%%%%%
\begin{figure}[t]
\centering
%\vspace{1cm}
\includegraphics[width=.47\textwidth,angle=0]{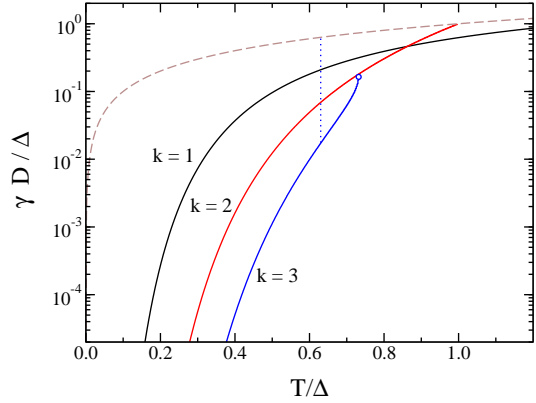}
\caption{Logarithm of the diffusivity $D$ as a function of the temperature $T$ for $k\!=\!1,2,3$. For $k\!=\!3$ the notation is the same as in Fig.~\ref{thermo1}. The dashed line is the value of $D$ in the paramagnetic phase, $D=T/\gamma$.}
\label{diffusion}
\end{figure}
%%%%%%%%%%%%%%%%%%%%%%%%%%%%%%%%%%

\noindent
The analytical expression for the diffusion constant of Eq.~(\ref{effdynresc}) is found for example in \cite{Risken} and is given by:
\begin{equation}
\tilde{D}(\tilde{T}) = \lim_{\tilde{t} \rightarrow \infty}  \frac{1}{2\tilde{t}} \langle | \varphi(\tilde{t})-\varphi(0) |^2 \rangle =
\frac{\tilde{T}} 
{I_0 (\tilde{T}^{-1})^2} \ .
\end{equation}
Then, from the definition of $D$ given by Eq.~(\ref{defdiff}) and from Eq.~(\ref{legameeffettivo}) one gets
\beq
\label{diff}
D(T)=\frac{k\Delta \zeta^{k-1}}{\gamma} \tilde{D}(T)=\frac{T}{\gamma} I_0(k \beta \Delta \zeta(T)^{k-1})^{-2} \ .
\eeq
In Fig. \ref{diffusion} we report $D$ 
as a function of temperature $T$ in a semilogarithmic scale.
At low temperature Eq.~(\ref{diff}) predicts an Arrhenius behavior:
\begin{equation}
D(T \ll \Delta) \simeq \frac{2\pi \Delta k e^{k-1}}{\gamma} e^{-\beta 2 \Delta k} \ .
\label{diff_low}
\end{equation}
In the high temperature paramagnetic phase one has $\zeta\equiv 0$ so that Eq.~(\ref{effdynresc}) reduces to a free Brownian motion and the diffusion constant is simply given by $D=T/\gamma$ \cite{nota2}. It is worth to note that for $k=1$ the paramagnetic phase does not exist and $D=T/\gamma$ is only the asymptotic limit of Eq.~(\ref{diff}) for $T \rightarrow \infty$.

%%%%%%%%%%%%%%%%%%%%%%%%%%%%%%%%%%%%%%%%%%%%%%%%%%%%%%%%%%%%%%%%%%%%%%%%%%%%%%%

\subsection{The self correlation function}

\noindent
As we already discussed, the model is not rotationally invariant, and when the $C_{kv}$ symmetry is broken a spontaneous magnetization appears, which phase can be an integer multiple of $2 \pi / k$. We will choose the phase to be zero in order not to break the $\varphi_i \rightarrow -\varphi_i$ symmetry. In this situation it is expected that even and odd functions of $\varphi$ have different behavior. Therefore, it is convenient to separate the contributions in $F(t)$ given by the real and imaginary part of $e^{i\varphi_i}$. Using the relations $\langle \cos(\varphi_i(t)) \sin(\varphi_i(0) \rangle = 0$ and $\langle \sin(\varphi_i(t)) \rangle=0$, due to the unbroken symmetry $\varphi_i \rightarrow -\varphi_i$, we can define from Eq.~(\ref{incoerente}) (setting $t'=0$ because of the time translation invariance):
\beq
\label{selfs}
\begin{split}
&F(t)=F_c(t) + F_s(t) \ , \\
&F_c(t)= \frac{1}{N} \sum_i \left[ \langle \cos \varphi_i(t) \cos\varphi_i(0) \rangle - \langle \cos \varphi_i(t) \rangle \langle \cos \varphi_i(0) \rangle \right] \ , \\
&F_s(t)= \frac{1}{N} \sum_i \langle \sin \varphi_i(t) \sin\varphi_i(0) \rangle  \ . \\
\end{split}
\eeq
As previously discussed, see  appendix \ref{app_self} or
Eq.~(\ref{legameeffettivo}), the above correlations are equal to the
ones calculated using the effective dynamical system (\ref{effdynsyst}); at equilibrium they are given by
\beq
\label{effcorr}
F(t)={\cal F}(t) \equiv  \langle e^{i[\varphi(t) - \varphi(0)]} \rangle_{\cal H} - \langle e^{i \varphi(t)} \rangle_{\cal H} \langle e^{-i \varphi(0)} \rangle_{\cal H}
\eeq
We will use for ${\cal F}(t)$ the same notation used for $F(t)$, see Eqs.~(\ref{selfs}).

\subsubsection{The reduced system}

\noindent
To compute the correlation functions, it is useful to use again the
reduced variables and Eq.~(\ref{effdynresc}).  Unfortunately, we have
not been able to derive an analytic expression for the correlations in
the whole $\tilde{T}$ range, but only in the high and
low temperature limits. In these limits the correlation functions
turns out to be exponentials: 
\beq
\begin{split}
\label{selfexp}
{\cal F}_c(\tilde{t})&= \langle \cos \varphi(\tilde{t}) \cos \varphi(0) \rangle - \langle \cos \varphi \rangle^2= A_c \exp \left[ -\frac{\tilde{t}}{\tilde{\tau}_c} \right] \ , \\
{\cal F}_s(\tilde{t})&=  \langle \sin \varphi(\tilde{t}) \sin \varphi(0) \rangle = A_s \exp \left[ -\frac{\tilde{t}}{\tilde{\tau}_s} \right] \ .
\end{split}
\eeq
%%%%%%%%%%%%% CORRSELF1 %%%%%%%%%%%%%%
\begin{figure}[t]
\centering
%\vspace{1cm}
\includegraphics[width=.75\textwidth,angle=0]{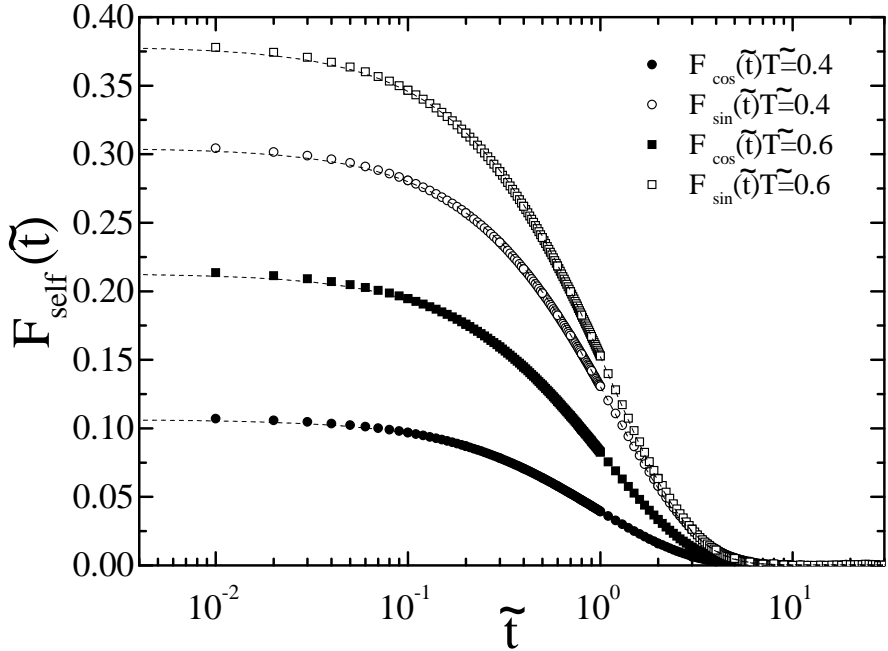}
\caption{The correlation function of $\cos \varphi$ and $\sin \varphi$ calculated using Eq.~(\ref{effdynresc}) for two different values of $\tilde{T}$. They are well fitted by an exponential form.}
\label{corrself1}
\end{figure}
%%%%%%%%%%%%%%%%%%%%%%%%%%%%%%%%%%%%%%%%%
It is important to note that the amplitudes of these correlations (which are equal to the ones of the original system as the variables $\varphi$ are not rescaled) are analytically computable at each temperature and are given by
\beq
\begin{split}
\label{Ampiezzeint}
A_c&=\langle \cos^2 \varphi \rangle - \langle \cos \varphi \rangle^2 = 1 - \tilde{T} \zeta(\tilde{T}) - \zeta(\tilde{T})^2 \ , \\
A_s&= \langle \sin^2 \varphi \rangle = \tilde{T} \zeta(\tilde{T}) \ .
\end{split}
\eeq where $\zeta(\tilde{T})$ is the magnetization expressed as a
function of $\tilde{T}$ and determined self-consistently by
Eq.~(\ref{zeta}).  The relaxation times are related to the real ones
by $\tau_{c,s}= \frac{\gamma}{k\Delta \zeta^{k-1}}
\tilde{\tau}_{c,s}$.  The limits in which Eq.~(\ref{selfexp}) are
analytically obtained are:
\begin{itemize}
\item{\it High $\tilde{T}$ limit}: if $\tilde{T} \rightarrow \infty$
one can neglect in Eq.~(\ref{effdyneq}) the term proportional to $\sin
\varphi$. In this case the dynamics is close to a free Brownian motion
and the correlation functions are exponentials with $\tilde{\tau}_c
\sim \tilde{\tau}_s \sim 1/\tilde{T}$ and $A_s \sim A_c \sim 1/2$.
\item{\it Low $\tilde{T}$ limit}: the low temperature limit is
obtained by considering $\varphi(\tilde{t}) \ll \pi/2$ and {\it i)} by
expanding $\sin \varphi \sim \varphi$ in Eq.~(\ref{effdynresc}) and {\it
ii)} by approximating ${\cal F}_s(\tilde{t}) \sim \langle
\varphi(\tilde{t}) \varphi(0) \rangle$ and ${\cal F}_c(\tilde{t}) \sim
\frac{1}{4} [ \langle \varphi^2(\tilde{t}) \varphi^2(0) \rangle -
\langle \varphi^2 \rangle^2 ]$. In the approximation {\it i)} the
equation of motion (\ref{effdynresc}) is easily solved, the
correlation functions are exponentials and one obtains 
\beq
\label{lowTexp}
\begin{split}
A_c&=\frac{\tilde{T}^2}{2} \ , \hspace{1cm} \tilde{\tau}_c = \frac{1}{2} \ , \\
A_s&=\tilde{T} \ ,  \hspace{1.3cm} \tilde{\tau}_s = 1 \ .
\end{split}
\eeq 
The expressions for the amplitudes are consistent with
Eq.~(\ref{Ampiezzeint}) observing that in this limit
$\zeta \sim 1-\frac{\tilde{T}}{2} - \frac{\tilde{T}^2}{8}$.
\end{itemize}
%%%%%%%%%%  CORRSELF2  %%%%%%%%%%%%
\begin{figure}[t]
\centering
%\vspace{1cm}
\includegraphics[width=.75\textwidth,angle=0]{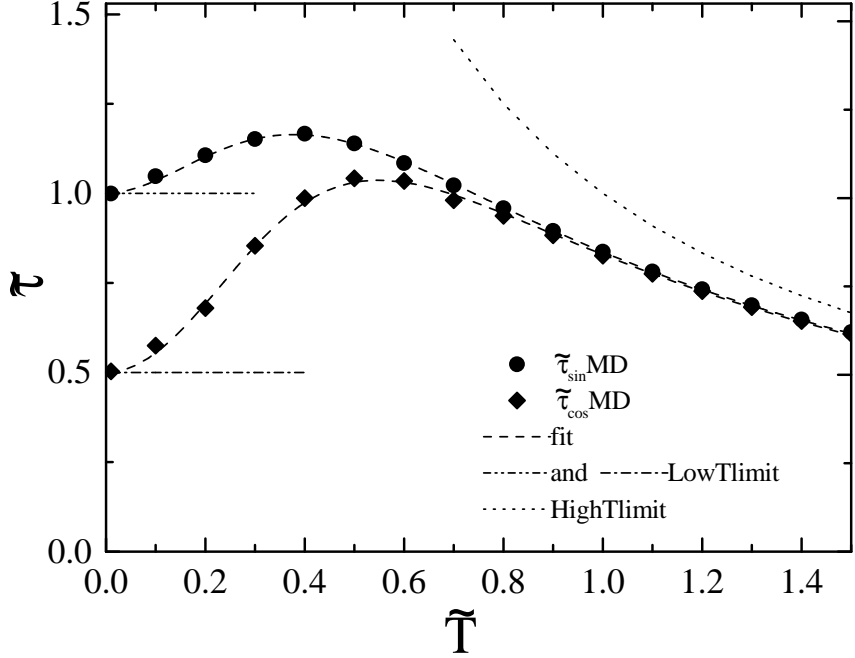}
\caption{Relaxation times of the correlation functions of Eq.~(\ref{effdynresc}). The symbols are the result of the numerical calculation, while the dashed lines are the fits and the dot (dot-dashed) lines is the low (high) temperature expansions.}
\label{corrself2}
\end{figure}
%%%%%%%%%%%%%%%%%%%%%%%%%%%%%%%%%
The complete $\tilde{T}$ dependence is obtained by solving
Eq.~(\ref{effdynresc}) numerically: the numerical solution has been
performed using {\it i)} the true dynamical system (defined through
Eq.~(\ref{hamiltonian}) and (\ref{dinamicalang})), and {\it ii)} the
effective one defined by Eq.~(\ref{effdynresc}).  In both cases the
reduced variables ($\tilde{t}$ and $\tilde{\eta}$) have been used.
The integration of the true dynamical system allowed to derive
numerically the self and the collective correlation functions (in each
case for both the $\sin \varphi$ and $\cos \varphi$ variables). The
self correlation functions, that for check have been compared to those
obtained trought the integration of the effective dynamical systems,
were fitted to an exponential decay to derive the parameters $A_c$,
$A_s$, $\tau_c$ and $\tau_s$.  The simulated systems is composed of
$N=1000$ degrees of freedom (in the case of the effective dynamical
systems, the $N$ independent degrees of freedom have been used to
collect statistical average over the initial conditions). The equation
of motion have been integrated by a simple constant stepsize ($d
\tilde{t} = 0.001$) Runge-Kutta method, and the RAND \cite{nota_sim}
fortran routine has been used to generate the gaussian noise
$\tilde{\eta}$. At each temperature we performed an equilibration of
$5 \cdot 10^6$ integration steps followed by $5 \cdot 10^6$ steps of
data collection. The time history of the variables $\varphi$ have been
stored and a multi-step circular buffer scheme has been employed to
calculate the appropriate correlation functions.  We found that for
any $\tilde{T}$ the correlations are well (but not exactly) described
by Eq.~(\ref{selfexp}); in Fig.~\ref{corrself1} we report some
correlations calculated numerically for intermediate $\tilde{T}$
values, together with an exponential fit.  The relaxation times
obtained numerically are reported in Fig. \ref{corrself2}. 
%%%%%%%%%%  CORRSELF3  %%%%%%%%%%%%
\begin{figure}[t]
\centering
%\vspace{1cm}
\includegraphics[width=.47\textwidth,angle=0]{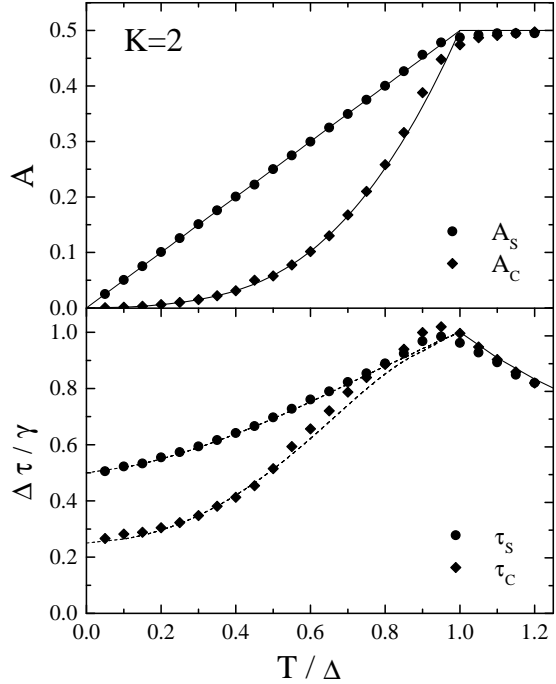}
\caption{The parameters of the self correlations for $k=2$. The symbols are the result of the numerical calculation. The full lines are obtained from analytic computation, while the dashed lines are obtained from the polinomial fit (Eq.~\ref{fittaccio}) on the reduced system and the substitution of $\tilde{T}$ with its value for $k=2$ (see text). The agreement is not perfect around $T_c$ due to finite size effects.}
\label{corrself3}
\end{figure}
%%%%%%%%%%%%%%%%%%%%%%%%%%%%%%%%%%
They have
been fitted for simplicity with a polinomial function: \beq
\label{fittaccio}
\tilde{\tau}(\tilde{T})= \frac{\tilde{\tau}(0)(1+P_1 \tilde{T}) + P_3
\tilde{T}^2}{1+P_1 \tilde{T} + P_2 \tilde{T}^2 + P_3 \tilde{T}^3} \ ,
\eeq 
where $\tilde{\tau}(0)$ is given by Eq.~(\ref{lowTexp}). The
previous expression reproduces the correct high and low $\tilde{T}$
limits. The values of the parameters $P_i$ are: 
\beq
\begin{split}
&\tilde{\tau}_c : \hspace{1cm} P_1=0  \hspace{1cm} P_2=0.90  \hspace{1cm} P_3=6.28 \\
&\tilde{\tau}_s : \hspace{1cm} P_1=1  \hspace{1cm} P_2=1.66  \hspace{1cm} P_3=6.28
\end{split}
\eeq
The relaxation times obtained by the numerical calculations are reported in Fig.~\ref{corrself2}
together with the corresponding fit and the high and low temperature expansions.
Having an (exact) expression for the amplitudes (Eq.~(\ref{Ampiezzeint})) and another one for the relaxation times (Eq.~(\ref{fittaccio}), extrapolated from numerical data) we can discuss the behavior of the correlation functions for any value of $k$ by substituting in these expressions $\tilde{T}=\frac{T}{k\Delta \zeta^{k-1}}$. 

\subsubsection{$k=1$}

\noindent
For $k=1$ we have $T=\Delta \tilde{T}$ and $\tau_{c,s}= \frac{\gamma}{\Delta} \tilde{\tau}_{c,s}$, so that the behavior of the relaxation times is obviously the same as in Fig. \ref{corrself2}. The amplitudes, that we do not report, are analytic functions of the temperature.

\subsubsection{$k=2$}

\noindent
As reported before, for $k=2$ a second order phase transition takes
place at $T_c=\Delta$. The parameters $A$ and $\tau$ for this case are
reported in Fig. \ref{corrself3}. We have that $\tilde{T} \rightarrow
\infty$ by approaching the phase transition from the magnetic phase,
so $A_{c,s} \rightarrow \frac{1}{2}$ and $\tau_{c,s} \rightarrow
\gamma/T_c = \gamma/\Delta$ at the transition. In the paramagnetic
phase one has $\zeta \equiv 0$, so that Eq.~(\ref{effdyneq}) reduces
to a free Browian motion; then $A_c=A_s=1/2$ and
$\tau_c=\tau_s=\gamma/T$ exactly for all temperatures above $T_c$. We
obtain then that $A$ and $\tau$ are continuous functions of
temperature but their derivatives have a discontinuity at $T_c$.

%%%%%%%%%%  CORRSELF4  %%%%%%%%%%%%
\begin{figure}[t]
\centering
%\vspace{1cm}
\includegraphics[width=.47\textwidth,angle=0]{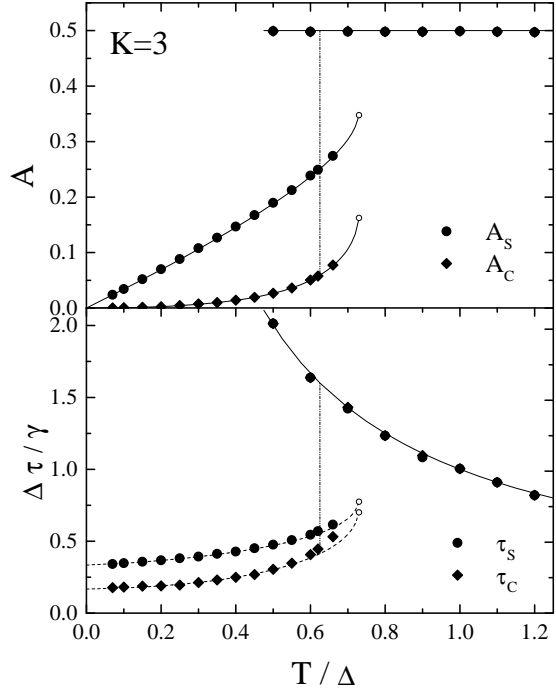}
\caption{The parameters of the self correlations for $k=3$ (with the same notations used in the previous figure). The vertical lines marks the transition temperature $T_0$; the numerical calculations can be done also in the metastable phases around $T_0$. The white dots mark the point at which the two solutions with $\zeta \neq 0$ disappear.}
\label{corrself4}
\end{figure}
%%%%%%%%%%%%%%%%%%%%%%%%%%%%%%%%%%

\subsubsection{$k \geq 3$}

\noindent
For $k \geq 3$ a first order phase transition takes place. The parameters $A$ and $\tau$ are reported in Fig. \ref{corrself4}. In this case $\tilde{T}$ does not diverge approaching the transition temperature, while the behavior in the high temperature paramagnetic phase is the same as for $k=2$. Then, obviously, $A$ and $\tau$ are discontinuous at the transition.

%%%%%%%%%%%%%%%%%%%%%%%%%%%%%%%%%%%%%%%%%%%%%%%%%%%%%%%%%%%%%%%%%%%%%%%%%%%%%%%%%%%%%%%%%%%%%%%%%%%%%%%%%%%%%%%%%%%%%%%%%%%%%%%%%%%%%%%%%%%%%%%%%%%%%%%%%%%%%%%%%%%%%%%%%%%%%%%%%%%%%%%%%%%%%%%%%%%%%%%%%%%%%%%%%

\section{Collective dynamics}
\label{sec:coll_dyn}

\begin{figure}[t]
\centering
%\vspace{1cm}
\includegraphics[width=.47\textwidth,angle=0]{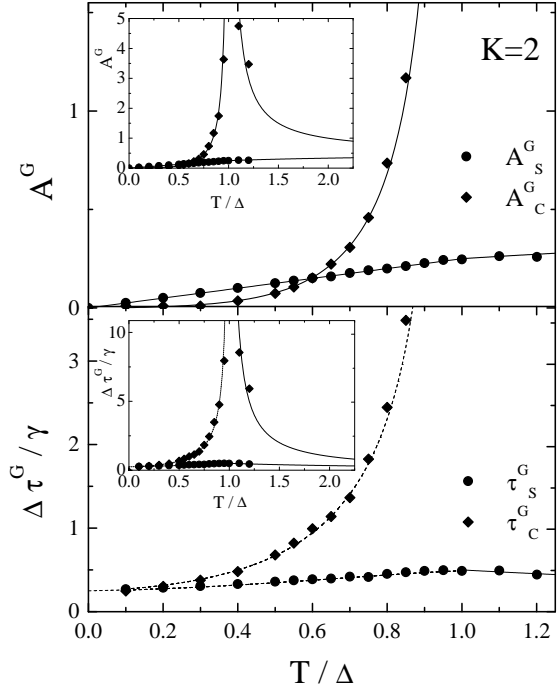}
\caption{The parameters of the collective correlations for $k=2$. As in the previous figures, the symbols are from numerical computation, the full lines are obtained analytically, while the dashed lines are obtained using the fitted expression (Eq.~(\ref{fittaccio})) in Eq.~(\ref{paramcollself}). In the inset the divergence at $T_c$ of the susceptibility and of the relaxation time related to the real part of the magnetization is evidenced.}
\label{corrcoll1}
\end{figure}

\begin{figure}[t]
\centering
%\vspace{1cm}
\includegraphics[width=.47\textwidth,angle=0]{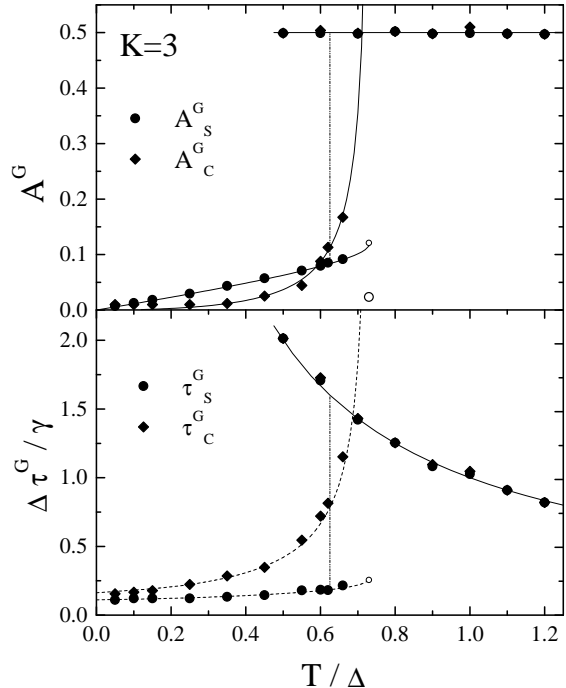}
\caption{The parameters of the collective correlations for $k=3$ (with the same notations used in the previous figure). In this case $A^G_c$ and $\tau^G_c$ diverge in the metastable phase when the metastable solution disappears (white dot in the figure).}
\label{corrcoll2}
\end{figure}

\noindent
To study the collective behavior of the system we introduce the correlation function of the magnetization $z$. It is defined by
\beq
G(t,t')=\frac{1}{N} \sum_{ij} \left[ \langle e^{i[\varphi_i(t)-\varphi_j(t')]} \rangle  -  \langle e^{i \varphi_i(t)} \rangle \langle e^{- i \varphi_j(t')} \rangle \right] = N \left[ \langle z(t) z^*(t') \rangle - \langle z(t) \rangle \langle z^*(t') \rangle \right] \ .
\eeq
Again, with the aim to obtain simple exponential behavior of the correlation functions, it is convenient to separate the contributions coming from the real and immaginary part of the magnetization, defining
\beq
\begin{split}
&G_c(t,t')=\frac{1}{N} \sum_{ij} \left[ \langle \cos \varphi_i(t) \cos \varphi_j(t') \rangle  -  \langle \cos \varphi_i(t) \rangle \langle \cos \varphi_j(t') \rangle \right] \ , \\
&G_s(t,t')=\frac{1}{N} \sum_{ij} \langle \sin \varphi_i(t) \sin \varphi_j(t') \rangle \ .
\end{split}
\eeq
These correlations are related to the Gaussian correction around the thermodynamic limit, i.e. the leading correction in $1/N$ for $N \rightarrow \infty$. In appendix \ref{app_coll} we derive a general expression (Eq.~(\ref{relgencollself})) that relates $G(t)$ to the self correlation function $F(t)$ defined in the previous section. Using this expression and assuming that {\it i)}~the magnetization is always real and {\it ii)}~$F_{c,s}(t)$ are given by Eq.~(\ref{selfexp}), $G_{c,s}(t)$ turn out to be also exponentials and are given by:
\beq
\label{GGG}
G_c(t)=Z_c A_c \exp \left[ -\frac{t}{Z_c \tau_c} \right] \ , \hspace{1cm} G_s(t)=Z_s A_s \exp \left[ -\frac{t}{Z_s \tau_s} \right] \ ,
\eeq
where
\beq
\label{ZETA}
Z_c=\frac{1}{1-\beta\Delta k(k-1)\zeta(\beta)^{k-2} A_c} \ , \hspace{1 cm} Z_s=\frac{1}{1+\beta\Delta k (k-1) \zeta(\beta)^{k-2} A_s} \ .
\eeq
As expected in absence of interactions, for $k=1$ we have $Z_c=Z_s=1$,
so that $G(t)=F(t)$ at all temperatures. \\
In order to treat the  $k \geq 2$ case we define
\beq
G_{c,s}(t)=A^G_{c,s} \exp \left[ -\frac{t}{\tau^G_{c,s}} \right] \ ,
\eeq
with
\beq
\label{paramcollself}
A^G_{c,s}=Z_{c,s} A_{c,s} \ , \hspace{1cm} \tau^G_{c,s}=Z_{c,s}
\tau_{c,s} \ .
\eeq
From the analytic expression for $A_{c,s}$, given
by Eq.~(\ref{Ampiezzeint}), we get an analytic expression for the
constants $Z_{c,s}$. Exact expression for the parameters $A_{c,s}^G$,
that are connected by the fluctuation-dissipation theorem
to the susceptibilities of the magnetization, can then be derived. The
relaxation times of $G(t)$ are obtained from the one of $F(t)$, that
we studied in the previous section, using Eq.~(\ref{fittaccio}). Then,
their expression is not exact but derives from the numerical data on
the reduced system that we defined in the previous section. To
emphasize this, in Fig. \ref{corrcoll1} and \ref{corrcoll2} we report
as a full line the exact expressions and with a dashed line the
expressions derived using the numerical solution of the reduced system
and Eq.~(\ref{paramcollself}).

%%%%%%%%%%  CORRCOLL1  %%%%%%%%%%%%
%%%%%%%%%%%%%%%%%%%%%%%%%%%%%%%%%%

%%%%%%%%%%  CORRCOLL2  %%%%%%%%%%%%
%%%%%%%%%%%%%%%%%%%%%%%%%%%%%%%%%%

\subsection{$k=2$}

\noindent
The parameters $A^G$ and $\tau^G$ given by Eq.~(\ref{paramcollself})
for $k=2$ are reported in Fig. \ref{corrcoll1}. From the symmetry
structure of the model, we know that it can acquire a spontaneous
magnetization in the directions $\psi=0$ and $\psi=\pi$, but not in
the orthogonal direction. Then we expect a divergence in the amplitude
and in the relaxation time of $G_c(t)$ but not in the same parameters
of $G_s(t)$. In the high temperature paramagnetic phase we have
$A_c=A_s=1/2$ and $\tau_c=\tau_s=\gamma/T$ (see the discussion of the
$k=2$ case in the previous section); then $Z_c=1/(1-\beta \Delta)$ and
$Z_s=1/(1+\beta \Delta)$.  From these expressions and
Eq.~(\ref{paramcollself}) we get an expression for $A^G$ and $\tau^G$
in the paramagnetic phase (full lines in Fig. \ref{corrcoll1}). It is
easy to see that close to $T_c$ one has $A^G_c \sim |T-T_c|^{-1}$ and
$\tau^G_c \sim |T - T_c|^{-1}$. The same behavior is obtained
approaching the transition temperature from below, as one can easily
check remembering that for $T \rightarrow T_c^-$ one has $\tilde{T}
\rightarrow \infty$, $A_{c,s} \rightarrow 1/2$, $\tau_{c,s}
\rightarrow \gamma/T$.  We obtain again the classical mean field
critical exponents for the universality class of our model. In the low
temperature phase, as previously discussed, the expression for the
relaxation times in not exact, and is reported as a dashed line in
Fig.~\ref{corrcoll1}.

\subsection{$k \geq 3$}

\noindent
The parameters $A^G$ and $\tau^G$ given by Eq.~(\ref{paramcollself}) for $k=3$ are reported in Fig. \ref{corrcoll2}. As previously shown, for $k \geq 3$ the model undergoes a first order transition at $T_0$.
The low temperature phase is metastable up to a certain temperature $T_c$ (see Fig. \ref{thermo1}) where it disappears, while the high temperature phase is metastable down to $T=0$.
In the high temperature phase we have, as in the $k=2$ case, $A_c=A_s=1/2$ and $\tau_c=\tau_s=\gamma/T$. But for $k \geq 3$ we have $Z_c=Z_s=1$, so that $G(t)=F(t)$.
We note that $A^G_{c,s}$ does not diverge in this phase, and the same happens to $\tau^G_{c,s}$ that diverge only for $T \rightarrow 0$.
In the low temperature phase it is easy to see (substituting Eq.~(\ref{Ampiezzeint}) in Eq.~(\ref{ZETA})) that $Z_s=1/k$; this happens for all $k$ if $\zeta \neq 0$.
Then, $A^G_s$ and $\tau^G_s$ are simply proportional to $A_s$ and $\tau_s$ respectively, and do not show any anomaly. The behavior of $Z_c$ is more interesting: using its definition given in Eq.~(\ref{ZETA}), Eq.~(\ref{zeta}) and Eq.~(\ref{Ampiezzeint}), one can show that $Z_c \rightarrow \infty$ when $T \rightarrow T_c^-$.
Thus, $A^G_c$ and $\tau^G_c$ diverge in the metastable region of the paramagnetic phase approaching the temperature at which the paramagnetic phase itself disappears. Again, the expression for $\tau^G$ is obtained using Eq.~(\ref{fittaccio}), and is reported as a dashed line in Fig.~\ref{corrcoll2}.

%%%%%%%%%%%%%%%%%%%%%%%%%%%%%%%%%%%%%%%%%%%%%%%%%%%%%%%%%%%%%%%%%%%%%%%%%%%%%%%%%%%%%%%%%%%%%%%%%%%%%%%%%%%%%%%%%%%%%%%%%%%%%%%%%%%%%%%%%%%%%%%%%%%%%%%%%%%%%%%%%%%%%%%%%%%%%%%%%%%%%%%%%%%%%%%%%%%%%%%%%%%%%%%%%%%

\section{Closest saddles to equilibrium configurations}
\label{selleW}

\noindent
The important role that stationary points (saddles) of the Potential
Energy Surface (PES) plays in the dynamics of various systems has been
clarified recently
\cite{noi_sad,cav_sad,sad_1,doye,sad_3,grig,sadBLJ,parisi_boson,cav2}.
From previous works it seems that, in order to describe the
equilibrium dynamics at a given temperature $T$, it is sufficient to
know the properties of some of them, that have often been called
``closest saddles to the equilibrium configurations at temperature
$T$'' \cite{parisi_boson2}.  To locate these particular stationary
points, two main strategies have been adopted: {\it 1)} defining in a
proper way a ``distance'' in phase space and, given an equilibrium
configuration, looking at the stationary point that has minimum
distance from this configuration; {\it 2)} partitioning the phase
space in ``basins of attraction'' of stationary points via an
appropriate function that has a local minimum on each stationary
point.  While the first approach has been exploited analytically on
some disordered spin models \cite{cav2}, the second one has been
extensively used in numerical simulations of simple model liquids
\cite{noi_sad,cav_sad,sad_1,doye,sad_3,grig,sadBLJ}. \\ 
The problem
with definition {\it 2)} is that one has to define a function $W$ such
that {\it each stationary point of $H$ is a local minimum of $W$} and
{\it each local minimum of $W$ is a stationary point of
$H$}. Otherwise, while looking for the closest saddle starting from a
reference configuration one can remain trapped in some local minimum
of $W$ that is not a stationary point of $H$. It has been shown in
\cite{doye} that this possibility effectively arises in the majority
of the cases if one chooses, as usual in simple liquids, $W=|\nabla
H|^2$. \\ To compare the two methods avoiding the difficulties of the
numerical computations, we tried to find some models in which the
minimization of a function $W$ with the desired property could be
analytically performed.  In the $k$-trigonometric model this function
can correctly be chosen as $W=|\nabla H|^2$, as one can check
directly: in fact, all the minima of $W$ correspond to stationary
points of $H$.  In this section we present a general method for the
minimization of $W=|\nabla H|^2$, that can be probably extended to
treat a large class of mean field systems without quenched disorder.
We apply this technique to our model and we show that definitions {\it 1)} and {\it
2)} give in this case {\it exactly} the same result.
Note that the idea on which the method is constructed can be
used also with $W$ functions different from the one chosen here, even
if the practical calculation might be difficult depending on the
particular form chosen for $W$.  Future work will be devoted to
apply our method, if possible, to disordered systems like the $p$ spin
disordered model studied in \cite{cav2}.

\subsection{Definition of the relevant quantities}

\noindent
In section \ref{topologia} we studied the properties of the stationary
points of the PES that are independent from the
statistical measure that describes the system at temperature $T$ (i.e. they are independent of temperature).
From this study we were able to find a relation between the energy of a saddle and its order and to compute
the number of saddles, $\exp N \sigma(e)$, located at a given energy $e$. Now we want to calculate the energy
of the ``closest saddles to equilibrium configurations at temperature $T$''.  The procedure used to calculate
this quantity is the same used {\it numerically} in \cite{noi_sad,cav_sad}: we consider an initial
configuration extracted from the Gibbs distribution at temperature $T$, and we perform a minimization of
\beq
\label{W}
W=|\nabla H|^2= \frac{N \Delta^2 k^2}{2} \left[ (z \ z^*)^{k-1} -
\Re(z^{2k-2} \ z_2) \right] \ , \hspace{1cm} z_2=\frac{1}{N} \sum_i
e^{2i\varphi_i} \ .  
\eeq 
that leads to the ``closest saddle to the
initial configuration''. Finally, we average over the equilibrium
distribution of initial data. \\ The minimization of $W$ is performed
using the dynamical system \beq
\label{minW}
\gamma \dot{\varphi}_i = - \frac{\partial W}{\partial \varphi_i} \ ,
\eeq 
that is completely analogous to (\ref{dinamicalang}) with
$H$$\rightarrow$$W$ and $T$$=$$0$. We want to calculate the energy of
the configurations in the limit of infinite time starting from a Gibbs
ensemble, i.e.  \beq
\label{esaddle}
e_s(T)=\frac{1}{N} \lim_{t \rightarrow \infty} \langle H (t) \rangle_W \ .
\eeq
From this quantity, we obtain the order of the saddles as a
function of temperature using Eq.~(\ref{x-v}) and the
``configurational entropy'' of the saddles that is given by $\sigma(T) =
\sigma(e_s(T))$.

\subsection{Effective dynamical system}

\noindent
Using the same argument presented in appendix \ref{app_self} for the
real dynamics, 
it is possible to show that the dynamical system (\ref{minW}) is equivalent to the single-particle one given by
\beq
\label{Wsingola}
\begin{cases}
{\cal W}(\varphi,t) = \Delta^2 k^2 (k-1) \ \Re \{ \ [ \ \zeta^*(t) \ (\zeta(t) \zeta^*(t))^{k-2} - \zeta_2(t) \ \zeta(t)^{2k-3} \ ] \ e^{i \varphi} \ \} - \frac{1}{2} \Delta^2 k^2 \ \Re \{ \ \zeta(t)^{2k-2} \ e^{2 i \varphi} \ \} \ , \\
\gamma \dot{\varphi}(t)=-\frac{\partial {\cal W}}{\partial \varphi}(\varphi(t),t) \ , \\
\zeta(t)=\langle e^{i \varphi(t)} \rangle \ , \\
\zeta_2(t)=\langle e^{2 i \varphi(t)} \rangle \ .
\end{cases}
\eeq
The derivation is sketched in appendix \ref{app_W}.  In this case
we will calculate the averages over the distribution of initial data
setting the noise to zero. As the system is mean field, the
correlations between different degrees of freedom vanish in the thermodynamic limit and the Gibbs
distribution can be written in the form 
\beq
\label{gibbs}
P(\{ \varphi \} ) = \prod_i {\cal P}( \varphi_i ) = \prod_i \frac{e^{ -\beta {\cal H}(\varphi_i)}}{{\cal Z}_i} = \prod_i \frac{e^{ \beta \Delta k \Re [ \zeta(\beta)^{k-1} e^{i \varphi_i}] }}{{\cal Z}_i} \ ,
\eeq
where $\zeta(\beta)=\zeta(t=0)$ is as usual the equilibrium average magnetization. The problem is then reduced to the calculation of 
\beq
\label{e_s_eff}
e_s(T)=\lim_{t \rightarrow \infty} \langle {\cal H}(\varphi) \rangle_{\cal W} \ ,
\eeq
using the dynamical system (\ref{Wsingola}) and averaging over the distribution (\ref{gibbs}) of initial data.

\subsection{Calculation of the energy of the closest saddles}

\noindent
Again we assume that the system is not magnetized or that the mean
magnetization has zero phase. So, we assume that $\zeta(t)$ and
$\zeta_2(t)$ are real functions of time. In this case the system
(\ref{Wsingola}) becomes 
\beq
\label{Wappoggio}
\begin{cases}
{\cal W}(\varphi,t) = \Delta^2 k^2 \ \zeta(t)^{2k-3} \ [ \ (k-1) \ (1
- \zeta_2(t) ) \ \cos \varphi - \frac{1}{2} \zeta(t) \cos 2 \varphi \
] \\ \gamma \dot{\varphi}(t)=-\frac{\partial {\cal W}}{\partial
\varphi}(\varphi(t),t) \\ \zeta(t)=\langle \cos \varphi(t) \rangle \\
\zeta_2(t)=\langle \cos 2 \varphi(t) \rangle
\end{cases}
\eeq
We want now to show that the first term in ${\cal W}$ can be
neglected, at least in some limits.  For $k=1$ the first term
disappears and one simply has ${\cal W}=-\frac{\Delta^2}{2} \cos 2
\varphi$. For $k \geq 2$, in the paramagnetic phase one has
$\zeta(0)=\zeta_2(0)=0$, then ${\cal W}=0$ and the closest saddle is
the starting configuration itself. In the low temperature phase we
know that at the initial time $1 - \zeta_2(0)= 2 \langle \sin^2
\varphi \rangle = \frac{2T}{k\Delta \zeta(0)^{k-2}}$. Then we can
neglect the first term with respect to the second one (at $t=0$) if
\beq (k-1) \frac{2T}{k\Delta \zeta(0)^{k-2}} \ll 4 \frac{1}{2}
\zeta(0) \hspace{1cm} \Leftrightarrow \hspace{1cm} \frac{T}{\Delta}
\frac{k-1}{k} \ll \zeta(0)^{k-1} \ , \eeq where the $4$ in the right
side comes from the fact that the derivative of the second term is
proportional to $2 \sin 2\varphi \sim 4 \varphi $ while the derivative
of the first term is proportional to $\sin \varphi \sim \varphi$. This
condition is clearly satisfied for low enough temperature because
$\zeta(T=0)=1$. It is easy to check (see Fig. \ref{thermo1}) that for
$k=3$ the inequality is satisfied up to the transition temperature
$T_0$. Obviously for $k=2$ it cannot be satisfied close to $T_c$ where
$\zeta \sim 0$. \\ If one can neglect the first term at $t=0$, it can
be neglected at all subsequent times, because during the minimization
of ${\cal W}$ both $\zeta(t)$ and $\zeta_2(t)$ increase, and $\zeta_2
\rightarrow 1$ rapidly. To give an argument, let us neglect again the
first term in ${\cal W}$; then the minima of the pseudopotential are
such that $\cos 2 \varphi = 1$, so that increasing time $\zeta_2(t)$
will move toward 1 that is its infinite time limit. $\zeta$ will
increase due to the fact that the final energy is lower than the
initial one, as we will show below. Surprisingly, while we expect all
these approximations to work only at low enough temperature, they give
the correct result in the whole temperature range, as we checked
numerically. \\ In the approximations discussed before, the system
(\ref{Wappoggio}) becomes of the form:
\begin{equation}
\left\{ 
\begin{array}{ll}
\dot{\varphi} = - \nabla {\cal{W}} = - \Delta^2 k^2 \zeta(t)^{2k-2} \sin 2\varphi \ , \\
\zeta(t) = \langle \cos \varphi(t) \rangle \ , \\
\end{array}
\right. 
\label{steep}
\end{equation}
%%%%%%%%%%  SADDLE2  %%%%%%%%%%%%
\begin{figure}[t]
\centering
%\vspace{1cm}
\includegraphics[width=.47\textwidth,angle=0]{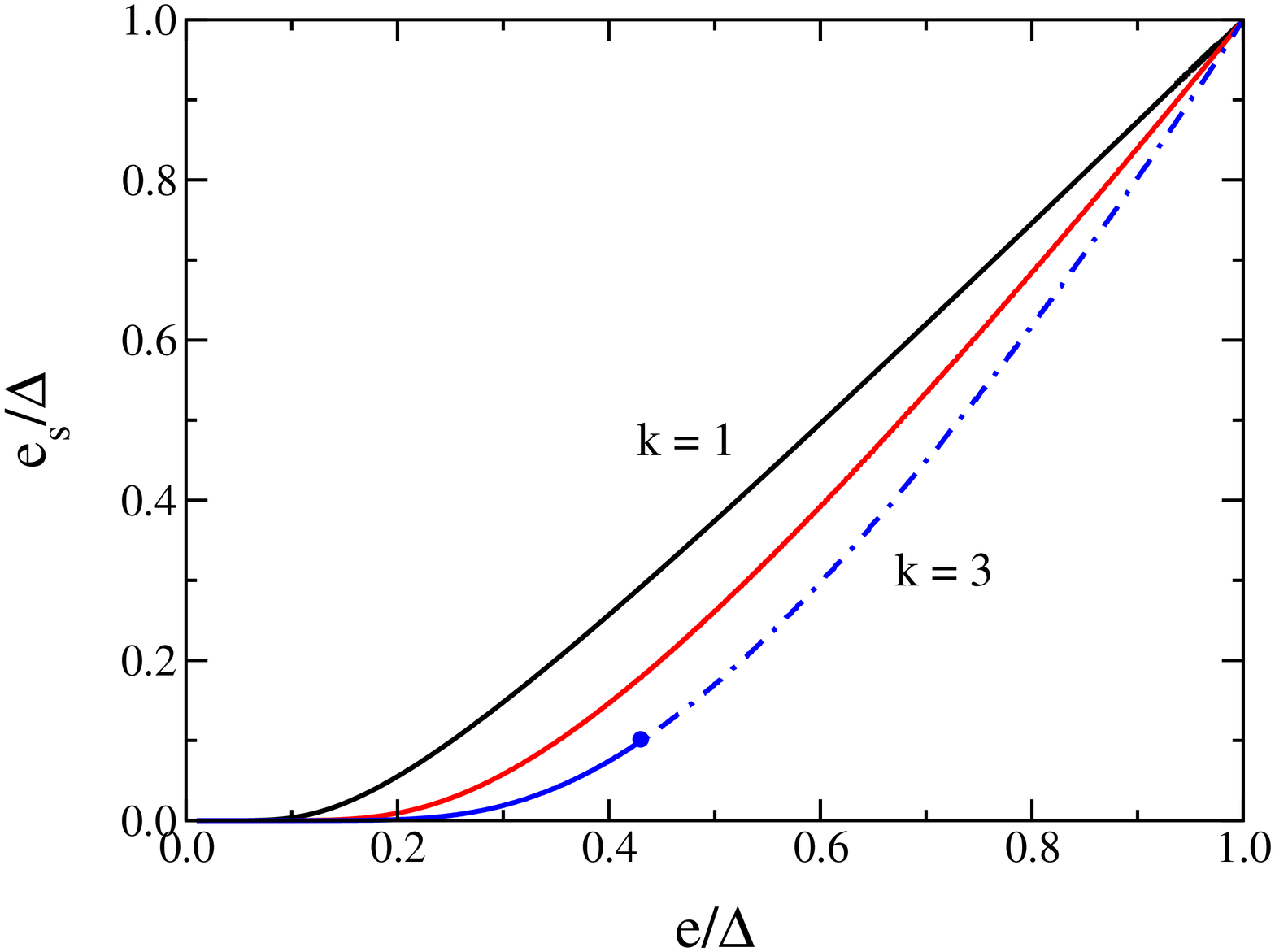}
\caption{Saddle energy $e_s$ versus the thermodynamic energy $e$ for $k=1,2,3$. The dashed line is the unstable region that corresponds to the solution of Eq.~\ref{zeta} that is a maximum of the free energy.}
\label{saddle2}
\end{figure}
%%%%%%%%%%%%%%%%%%%%%%%%%%%%%%%%%%
We are interested in the infinite time solutions of Eq.~(\ref{steep}),
$\varphi(t\!\rightarrow\!\infty|\varphi_0)$,
as a function of the initial conditions 
$\varphi_0 \equiv \varphi(t\!=\!0)$.
Without explicitly solving Eq.~(\ref{steep}), 
we observe that the sign of $\nabla {\cal{W}}$ at fixed 
$\varphi$ does not change during time, due to the fact that the time dependent factor 
in Eq.~(\ref{steep})
is always positive.
This implies that the specific time dependence of $\zeta$ does not affect the final point 
$\varphi(t\!\rightarrow\!\infty|\varphi_0)$
reached from a given initial condition,
rather it controls the rapidity of approaching this final point. 
It it easy to see that the solutions are:
\begin{equation}
\varphi(t\!\rightarrow\!\infty|\varphi_0) = 
\left\{
\begin{array}{ll}
0 & \hspace{.5cm} \mbox{ if \ \  $\varphi_0 \notin (\frac{\pi}{2},\frac{3\pi}{2})$ } \ , \\ \\
\pi & \hspace{.5cm} \mbox{ if \ \  $\varphi_0 \in (\frac{\pi}{2},\frac{3\pi}{2})$ } \ .
\end{array}
\right.
\end{equation}
The solution $\varphi \!=\!0$ is a minimum of ${\cal{W}}$ corresponding to 
a minimum of the effective potential energy (Eq.~(\ref{Heff})), 
while the solution $\varphi \!=\!\pi$ is a minimum of ${\cal{W}}$ that 
corresponds to a maximum of the effective potential.
The energy $e_s$ of the saddle is then obtained as $e_s(T)=\Delta (1 - \zeta(\infty)^k)$, where
$\zeta(\infty)$ is the average of $\cos \varphi(t\!\rightarrow\!\infty|\varphi_0)$ over the distribution
(\ref{gibbs}) of initial data:
\beq
\zeta(\infty) = \int_0^{2\pi} d\varphi_0 \ {\cal P}(\varphi_0) \ \cos \varphi(t\!\rightarrow\!\infty|\varphi_0) =  \int_{0}^{2 \pi} d\varphi_0 \ {\cal P}(\varphi_0) \ \text{sgn} (\cos \varphi_0) =
\frac{L_0(\beta \Delta k \zeta^{k-1})}{I_0(\beta \Delta k \zeta^{k-1})} \ ,
\end{equation}
where we have introduced the modified Struve function of order $0$:
$L_0 (\alpha) = 2 \pi^{-1} \int_0^{\pi/2}  d\varphi \sinh (\alpha \cos \varphi)$.
In Fig. \ref{thermo2} the saddle energies $e_s$ (dashed lines) are reported as a function of temperature
for $k\!=\!1,2,3$: qualitatively $e_s$ reproduces the shape of the potential energy $e$, 
and it is always below $e$, but coincides with $e$ in the paramagnetic region for $k\!\geq\!2$, as previously discussed.
%%%%%%%%%%  SADDLE1  %%%%%%%%%%%%
\begin{figure}[t]
\centering
%\vspace{1cm}
\includegraphics[width=.47\textwidth,angle=0]{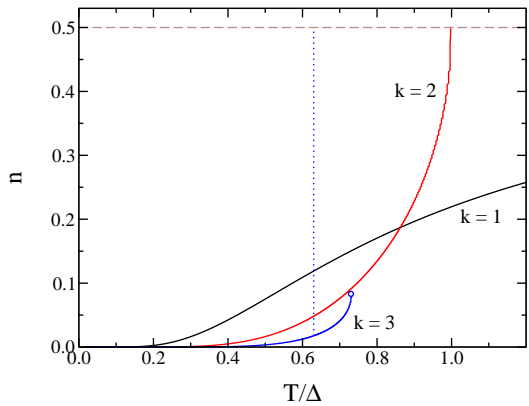}
\caption{Saddle order as a function of temperature for $k=1,2,3$.}
\label{saddle1}
\end{figure}
%%%%%%%%%%%%%%%%%%%%%%%%%%%%%%%%%%
The map $e_s$ vs. $e$ is shown in Fig.~\ref{saddle2}, where one
observes that, when the parameter $k$ increases,
the energy difference between instantaneous configurations and saddles becomes more and more pronounced.
From Eq.~(\ref{x-v}) we obtain the saddle order $n$ as a function of temperature:
\beq
\label{ord_sad}
n(T) = \frac{1}{2} \left[ 1 - \frac{L_0(\beta \Delta k \zeta^{k-1})}{I_0(\beta \Delta k \zeta^{k-1})} \right] \ .
\eeq
This function is reported in 
Fig. \ref{saddle1} for $k\!=\!1,2,3$.
At low temperature (high $\beta$) Eq.~(\ref{ord_sad}) is approximated by
\begin{equation}
n(\beta \gg 1) \simeq \sqrt{\frac{2 e^{k-1}}{\pi k \beta \Delta}} e^{-\beta \Delta k} \ ,
\label{ord_sad_low}
\end{equation}
which corresponds to an Arrhenius behavior.
We have shown in \cite{noidinamica} that the behavior of $n(T)$ is related to that of the diffusion
constant $D(T)$, as previously found numerically in simple model liquids \cite{noi_sad}, and that the energy barrier appearing in the Arrhenius low temperature expansion of $D(T)$ is exactly the energy
difference between saddles of order $1$ and the underlying minima.

\subsection{Distance of the closest saddle to the reference configuration}
\label{Wvsdistance}

\noindent
In this section we show that the two possible definitions of ``closest
saddles to equilibrium configurations'' that we discussed above are
coincide in our model. To this end, we apply the method introduced
in \cite{cav2} to our model. We compute the quantity
\beq \sigma(T;e_s,d)=\frac{1}{N} \int d\varphi_i \frac{e^{-\beta
H(\varphi)}}{Z(T)} \log \int d \psi_i \ \delta(H(\psi)-Ne_s) \
\delta(\partial_i H(\psi)) \ \det \text{H}(\psi) \ \delta \left( d -
d(\varphi,\psi) \right) \ , \eeq where $\text{H}_{ij}=\partial_i
\partial_j H$ is the Hessian matrix and $d(\varphi,\psi)$ is some
distance function between the two configurations $\varphi_i$ and
$\psi_i$. The argument of the logarithm is the number of stationary
points of energy $e_s$ and distance $d$ from the reference
configuration $\varphi$ (see appendix A or Ref. \cite{cav2} for a
detailed discussion). Then the logarithm of this number (divided by
$N$) is averaged over the equilibrium distribution at temperature $T$
of the reference configuration. \\ Using this quantity we can provide
a definition of ``closest saddles to equilibrium configurations''
\cite{cav2}: in fact, let the temperature be fixed (and neglect the
explicit dependence on it of all the quantities) and consider
$\sigma(e_s,d)$ as a function of $d$ at fixed $e_s$. This is the
number of saddles of energy $e_s$ and average distance $d$ from the
typical configurations at temperature $T$. We expect that for too
small distances this quantity will be zero, because there will be no
saddles of energy $e_s$ at too small distance from the equilibrium
configurations. So we can define $\bar{d}(e_s)$ as the value of $d$ at
which $\sigma(e_s,d)$ goes to zero: $\sigma(e_s,\bar{d}(e_s)) \equiv
0$. Then $\bar{d}(e_s)$ is the minimum distance at which one can find
saddles of energy $e_s$. Now we can minimize $\bar{d}(e_s)$ with
respect to $e_s$: the value $\bar{e}_s$ of $e_s$ such that
$\bar{d}(e_s)$ is minimum will be the energy of the closest saddles to
the equilibrium configurations, while $\bar{d}(\bar{e}_s)$ will be the
average distance from these saddles and the equilibrium configuration
themselves. \\
%%%%%%%%%%  OVERLAP  %%%%%%%%%%%%
\begin{figure}[t]
\centering
%\vspace{1cm}
\includegraphics[width=.47\textwidth,angle=0]{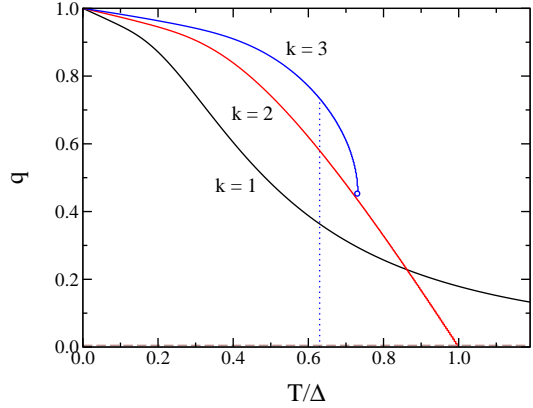}
\caption{Mean overlap between instantaneous configuration and the closest saddles (see text) as a function of temperature for $k=1,2,3$.}
\label{overlap}
\end{figure}
%%%%%%%%%%%%%%%%%%%%%%%%%%%%%%%%%%
In our model the distance function can be defined as
\beq d^2(\varphi,\psi)=1-q(\varphi,\psi)=1-\frac{1}{N} \sum_i
\cos(\varphi_i-\psi_i) \ .  \eeq In fact $\cos(\varphi_i-\psi_i)$ can
be interpreted as the scalar product of the unitary spins represented
by the angles $\varphi_i$ and $\psi_i$, so that $q(\varphi,\psi)$ is
the overlap between the two configurations. The calculations are
reported in appendix \ref{app_distance}; the result is that the energy
of the saddles is given by the same expression derived in the previous
section: \beq \bar{e}_s=e_s(T)=\Delta \left[ 1 - \left(
\frac{L_0(\beta \Delta k \zeta^{k-1})}{I_0(\beta \Delta k
\zeta^{k-1})} \right)^k \right] \ , \eeq where $\zeta$ is given by the
thermodynamics, see Eq.~(\ref{zeta}), while the mean overlap is given by
\beq \bar{q}= \int_0^{2\pi} d\varphi \ {\cal P}(\varphi) \ | \cos
\varphi |=\langle | \cos \varphi | \rangle_{\cal H} = \frac{L_1(\beta
\Delta k \zeta^{k-1})}{I_0(\beta \Delta k \zeta^{k-1})} \eeq where
${\cal P}$ has been defined in Eq.~(\ref{gibbs}) and
$L_1(\alpha)=L_0'(\alpha)$. Note that this result could also be
derived using the method of the previous section: in fact, we would
have, in analogy to Eq.~(\ref{e_s_eff}):
\beq
 q_s(T) = \lim_{t
\rightarrow \infty} \langle \cos ( \varphi(t) - \varphi_0)
\rangle_{\cal W} = \langle \cos ( \varphi(t \rightarrow \infty |
\varphi_0) - \varphi_0 ) \rangle_{{\cal H}(\varphi_0)} = \langle |
\cos(\varphi) | \rangle_{\cal H} = \bar{q}(T) \ . 
\eeq 
Then, we can conclude
that in our model the minimization of $W$ starting from an instantaneous configuration
equilibrated at temperature $T$ leads, on average, to that stationary point of the PES which has
maximum overlap with the starting configuration itself.
Moreover, we can calculate the average overlap (or
distance) between the equilibrium configurations and the associated stationary points,
that is reported in Fig.~\ref{overlap} as a function of
temperature for $k=1,2,3$.

\section{Conclusions}

\noindent
In this paper we presented a study of the thermodynamics and the
dynamics of a very simple mean field model of $N$ variables
interacting trough a fully-connected $k$-body trigonometric
term. In spite of its simplicity this 
model shows an interesting behavior undergoing second or first
order phase transitions depending on the value of $k$.  All the
results that we present here have been  obtained analytically
(except for the computation of the relaxation times of the self
correlation function that has been done numerically).  In particular,
the dynamics is analyzed in detail, and we find a relation between the
single-particle dynamics and the collective one in terms of a
Dyson-like equation that links the self and collective correlation
functions. Using this relation we studied, among other dynamical 
properties,  the critical
slowing down around the second order phase transition or close to the
stability limit in the case of the first order phase transition.  A
rather complete characterization of the geometry of the Potential Energy
Surface (PES) has been achieved. This allowed us  to relate some of the macroscopic
properties of the model to the PES characteristics: the
thermodynamic singularities (phase transitions) are located at the
same energy of the geometrical  ones, and, in the dynamics, the
low-temperature behavior of the diffusion constant is similar to the
behavior of the order of the saddles visited during the equilibrium
dynamics.  The concept of ``saddles visited during the equilibrium
dynamics'' (or ``generalized inherent structures'') has been widely
used in the literature, even if an unambiguous definition of them has
not yet been found. In this respect, we compared two 
definitions that have been used in the past, and we found that they give
exaclty the same result when applied to the $k$TM.  
This result supports the use of PES to analyze the behavior of 
interacting systems and suggests that the present analysis could be
applied to other interesting systems.

\acknowledgments

It is a pleasure to thank Andrea Cavagna for some
very illuminating discussions about the contents of sections
\ref{selleW}.

\vspace{0.2cm}

L. F. C. is research associate at ICTP Trieste and a fellow of the
Guggenheim Foundation. This reasearch was supported in part by
the National Science Foundation under Grant No. PHY99-07949
and the ACI Jeunes Chercheurs ``Algorithmes
d'optimisation et syst\`emes desordonn\'es quantiques''.

%%%%%%%%%%%%%%%%%%%%%%%%%%%%%%%%%%%%%%%%%%%%%%%%%%%%%%%%%%%%%%%%%%%%%%%%%%%%%%%%%%%%%%%%%%%%%%%%%%%%%%%%%%%%%%%%%%%%%%%%%%%%%%%%%%%%%%%%%%%%%%%%%%%%%%%%%%%%%

\newpage

\appendix

\section{Topological properties of the saddles in mean field models}
\label{app_euler}

\noindent
In this appendix we will calculate the number of stationary points of
energy $e=E/N$ and their order using a general method that works well
for all mean field models. Although this method is well known
\cite{cav2,chi_e_SUSY} and in our model the same results can be
obtained in a simpler way, it is useful to recall it here because it
will be generalized in appendix \ref{app_distance} to calculate the
number of saddles of energy $e$ located at a given distance from a
reference configuration equilibrated at a given temperature $T$. \\ We
introduce the quantity
\begin{equation}
\label{chi_e_def}
\chi(E) \equiv \int d \varphi_i \ \delta(H-E) \ \delta(\partial_i H) \ \det \text{H}= \sum_{\nu=0}^N (-1)^\nu {\cal N}_\nu(E)
\end{equation}
where H is the Hessian matrix ($\text{H}_{ij} = \partial_i \partial_j
H$), ${\cal N}_\nu(E)$ is the number of stationary points of $H$ of
order $\nu$ and energy $E$, and a product over the index $i$ is
omitted. The last equality is easily checked by noting that the
function $\det \text{H} \ \prod_i \delta(\partial_i H)$ is equal to
$(-1)^\nu$ if integrated in a small volume around a stationary point
of order $\nu$.  Using the relations
\begin{equation}
\begin{split}
&\det \text{H} = \int  d\bar{\eta}_i d\eta_i \ e^{ \sum_{ij} \eta_i \text{H}_{ij} \bar{\eta}_j} \\
&\delta(\partial_i H)= \int d\lambda_i \ e^{2\pi i\lambda_i \partial_i H}
\end{split}
\end{equation}
where $\{\eta,\bar{\eta}\}$ are Grassman variables \cite{ZJ} we have
\begin{equation}
\chi(E)=\int \frac{d \beta}{2 \pi} e^{\beta E} \ \int  d \varphi_i  d \bar{\eta}_i   d \eta_i  d \lambda_i \ \exp \left[ -\beta H + \eta_i \ \partial_i \partial_j H \ \bar{\eta}_j + 2 \pi i \lambda_i \partial_i H \right]
\end{equation}
Introducing the superfield
\begin{equation}
\label{superfield}
\phi_i(\theta,\bar{\theta})=\varphi_i + \eta_i \bar{\theta} + \bar{\eta}_i \theta + 2\pi i \lambda_i \theta \bar{\theta}
\end{equation}
where $\theta$ and $\bar{\theta}$ are two other Grassman variables we have
\begin{equation}
\label{chi_E_part}
\chi(E)=\int \frac{d \beta}{2\pi} e^{\beta E} \ \int  {\cal D}\phi_i \exp \left[ \int d\bar{\theta} d\theta (1-\beta \theta \bar{\theta}) H(\phi) \right]
\end{equation}
The last equality is easily checked remembering that
\begin{equation}
H(\phi)=H(\varphi) + (\eta_i \bar{\theta} + \bar{\eta}_i \theta + i \lambda_i \theta \bar{\theta})  \partial_i H(\varphi) + (\eta_i \bar{\eta}_j \theta \bar{\theta}) \partial_i \partial_j H(\varphi)
\end{equation}
due to the fact that $\theta^2=\bar{\theta}^2=0$. In mean field models we can evaluate the integral (\ref{chi_E_part}) at the saddle-point; form Eq.~(\ref{chi_e_def}) we see that the integral will be dominated by a particular value of $\nu$:
\begin{equation}
\chi(E) \sim_{N \rightarrow \infty} (-1)^{\bar{\nu}(E)} {\cal N}_{\bar{\nu}(E)}(E) \equiv e^{N \sigma(e)}
\end{equation}
so that
\begin{equation}
\lim_{N \rightarrow \infty} \frac{1}{N}\log \chi(E) = \lim_{N \rightarrow \infty} \frac{1}{N} \Big( \log {\cal N}_{\bar{\nu}(E)}(E) + i \pi \bar{\nu}(E) \Big)
\end{equation}
Then we expect that at the saddle point the real part of $\sigma(e)$ will be the logarithm of the number of saddles located at energy $e$, while its imaginary part will be the order of these saddles \cite{cav2}. We will now calculate explicitly all this quantities in our model. \\
The hamiltonian of the $k$-trigonometric model is written in term of the the variable $z=N^{-1} \sum_i e^{i\varphi_i}$ in Eq.~(\ref{hamiltonian}). This variable has a real and an imaginary part. As we want to evaluate Eq.~(\ref{chi_E_part}) at the saddle point in $z$, we will need to consider the real and imaginary part of $z$ as complex variables themselves. To avoid confusion, it is convenient to use another imaginary unit $I$ and $z=N^{-1}\sum_i e^{I \varphi_i} = \Re z + I \Im z$. When we will consider $\Re z$ and $\Im z$ as complex numbers themselves, we will use the notation $\Re z = \text{Re} \Re z + i \text{Im} \Re z$.
Setting $\Delta=1$ we have (neglecting all the constant prefactors)
\begin{equation}
\begin{split}
\chi(e)=&\int d \beta e^{\beta E} \ \int {\cal D}z \ \delta \left( N z- \sum_i e^{I \phi_i} \right) \ \int  {\cal D}\phi_i \exp \left[ \int d\bar{\theta} d\theta (1-\beta \theta \bar{\theta}) N (1 - \Re z^k) \right] \\
=&\int d\beta e^{N\beta e} \int {\cal D}z {\cal D}\hat{z} \exp N \left\{ \int d\bar{\theta} d\theta  \left[ (1-\beta \theta \bar{\theta}) (1 - \Re z^k ) +  \Re ( z \ i \hat{z}) \right] + \log {\cal X} (i\hat{z}) \right\} \\
{\cal X} (i\hat{z}) =& \int {\cal D}\phi \ \exp \left[ - \int d\bar{\theta} d\theta \ \Re (e^{I\phi} \ i \hat{z}) \right] 
\end{split}
\end{equation}
where we introduced the supervariables $z=z_0 + z_1 \theta + z_2 \bar{\theta} + z_3 \theta \bar{\theta}$ and $\hat{z}$ and the superdelta-function
\begin{equation}
\delta (z) = \int {\cal D} \hat{z} \ \exp \left[ \int d\bar{\theta} d\theta \ \Re (z \ i\hat{z}) \right]
\end{equation}
In the definition of $\delta(z)$ and of ${\cal X}(i\hat{z})$ we note that both imaginary units appear: $I$ serves to select only the component $\Re \hat{z} \cos \phi - \Im \hat{z} \sin \phi$ in the product $e^{I \phi} \hat{z}$, but $\Re \hat{z}$, $\Im \hat{z}$ and $\phi$ are themselves complex superfunctions with respect to $i$. We will also rotate the integration path on $\Re \hat{z}$ and $\Im \hat{z}$ in the complex ($i$) plane (that is equivalent to the substitution $i\hat{z} \rightarrow \hat{z}$); this is irrelevant because at the end we will look for the saddle point in the whole complex ($i$) plane.
We obtain then
\begin{equation}
\begin{split}
&\chi(e) = \int d\beta {\cal D}z {\cal D}\hat{z} \ e^{N \sigma(z,\hat{z},\beta \ | \ e)} \\
&\sigma(z,\hat{z},\beta \ | \ e)=\beta e +  \int d\bar{\theta} d\theta  \left[ (1-\beta \theta \bar{\theta}) (1 - \Re z^k ) + \Re z \hat{z} \right] + \log {\cal X} (\hat{z})
\end{split}
\end{equation}
To solve the saddle-point equations we will assume that at the saddle-point: {\it i)} all the fermionic components vanish ($z_1=z_2=\hat{z}_1=\hat{z}_2=0$) and {\it ii)} the $\Im$ part of the bosonic components is always 0 ($\Im z_0=\Im z_3=\Im \hat{z}_0 = \Im \hat{z}_3=0$). The first assumption is standard in this kind of computations. The second one is a consequence of the symmetry structure of the model: one can always choose the magnetization $z=N^{-1} \sum_i e^{I \varphi_i}$ such that its imaginary (in the $I$ plane) component is zero.
Firstly, we will evaluate ${\cal X}(\hat{z})$ in the case in which $\hat{z}$ has the form that we have assumed above.
We get
\beq
\begin{split}
{\cal X}(\hat{z}) &= \int {\cal D}\phi \ e^{ - \int d\bar{\theta} d\theta (\hat{z}_0 + \hat{z}_3 \theta \bar{\theta}) \cos \phi} = \int d\varphi \ d\bar{\eta} \ d\eta \ d\lambda \ e^{\hat{z}_0 (2 \pi i \lambda \sin \varphi + \eta \bar{\eta} \cos \varphi) - \hat{z}_3 \cos \varphi} \\
&= \int_0^{2 \pi} d\varphi \ e^{-\hat{z}_3 \cos \varphi} \ \delta(\hat{z}_0 \sin \varphi) \hat{z}_0 \cos \varphi
\end{split}
\eeq
If $\hat{z}_0$ is real and different from 0 we get
\begin{equation}
\label{i0el1}
{\cal I}_0 (\hat{z}_3) = -2 \ \text{sgn} \hat{z}_0 \ \sinh \hat{z}_3 
\end{equation}
Then we obtain
\beq
\label{sigma_top}
\sigma(z,\hat{z},\beta \ | \ e)=\beta e +  z_0 \hat{z}_3 + z_3 \hat{z}_0 - \beta (1-z_0^k) - k z_0^{k-1} z_3 + \log (-2  \ \text{sgn} \hat{z}_0 \ \sinh \hat{z}_3)
\eeq
The saddle-point equations are then
\begin{equation}
\begin{cases}
 e = 1 - z_0^k \\
 \hat{z}_0 = k z_0^{k-1}  \\
 \hat{z}_3 = - \beta k z_0^{k-1} \\
 z_0 = - \frac{1}{\tanh \hat{z}_3} \\
z_3 = 0
\end{cases}
\end{equation}
Substituting in $\sigma$ we obtain
\begin{equation}
\sigma(e)=z_0 \hat{z}_3 + \log 2 \sinh \hat{z}_3 + \log \text{sgn} \hat{z}_0
\end{equation}
Now, if $k$ is odd or if $k$ is even and $e < 1$, $z_0$ is real and given by $z_0 = (1-e)^{1/k}$. Then $\hat{z}_0$ is real and Eq.~(\ref{i0el1}) is correct.
Note also that $\hat{z}_0$ is positive (because $k-1$ is even or $e<1$) so that the last term in $\sigma$ is 0. Recalling that $\text{atanh} x = \frac{1}{2} \log \frac{1+x}{1-x}$ one has
\beq
\label{hatz3eul}
\hat{z}_3 = \text{atanh} \left( -\frac{1}{z_0} \right) =  \frac{1}{2} \log \frac{1-z_0}{1+z_0} +   \frac{i \pi}{2} 
\eeq
It is interesting to note that as $\hat{z}_3$ is complex while $z_0$ is real, form the third of the saddle point equations one obtains that $\beta$ is complex at the saddle point; this is a consequence of the srongly oscillating behavior of $\chi(e)$.
Using the relation
$\sinh x = \frac{ \tanh x}{\sqrt{1 - \tanh^2 x}}$
we get
$\sinh \hat{z}_3 = \frac{i}{\sqrt{1-z_0^2}}$
and introducing the variable 
\begin{equation}
n(e)= \frac{1}{2} (1-z_0(e)) =  \frac{1}{2} \left[ 1-(1-e)^{1/k} \right]
\end{equation}
we finally obtain
\begin{equation}
\sigma(n)=-n \log n - (1-n) \log (1-n) - i \pi n
\end{equation}
and
\beq
\chi(n(e)) \sim (-1)^{Nn(e)} e^{N \sigma(n(e))}
\eeq
This result is consistent with the one obtained in section \ref{topologia} and with the discussion at the beginning of this appendix. The case in which $k$ is even and $e > 1$ is a little more involved and we will not discuss it here.

%%%%%%%%%%%%%%%%%%%%%%%%%%%%%%%%%%%%%%%%%%%%%%%%%%%%%%%%%%%%%%%%%%%%%%%%%%%%%%%%%%%%%%%%%%%%%%%%%%%%%%%%%%%%%%%

\section{Single-particle dynamics}
\label{app_self}

\noindent
In this appendix we will show that one can use the effective dynamical system (\ref{effdynsyst}) to compute one particle quantities (see Eq.~(\ref{onepartobs})). We will restrict to the observable $e^{i\varphi(t)}$ and its n-times correlations; the other observables are linear combinations of this one (via a Fourier expansion).
We will use the formalism of the generating functional in its
supersymmetric formulation as presented 
in \cite{ZJ,cugliandolo1,cugliandolo2}, and a notation similar to the one of appendix A.

\subsection{The generating functional}

\noindent
The generating functional of the correlation functions can be written as \cite{ZJ,cugliandolo2,cugliandolo1}
\beq
\label{genfunct}
Z[h(t)]=\int {\cal D}\phi_i \ \exp \left[ \ \frac{1}{2} \sum_i \int \  da \ \phi_i(a) \ \Delta^{(2)} \ \phi_i(a) - \int da \ H(\phi) + \Re \int \ da \ h(a) \ e^{I \phi_1} \ \right]
\eeq
where $\theta$, $\bar{\theta}$ are Grassman variables, $\phi_i(\theta,\bar{\theta},t)$ is a time-dependent superfield, see Eq.~(\ref{superfield}), $da=d\bar{\theta}d\theta dt$, $h(a)=h(t) \theta \bar{\theta}$, $h(t)=\Re h(t) + I \Im h(t)$ and
\beq
\Delta^{(2)}=2T\frac{\partial^2}{\partial \bar{\theta} \partial \theta} - 2 \gamma \theta \frac{\partial}{\partial \theta}\frac{\partial}{\partial t} + 4 \gamma \theta \bar{\theta} \frac{\partial^2}{\partial \bar{\theta} \partial \theta} \frac{\partial}{\partial t}
\eeq
In fact it is easy to check that the self-correlation functions (\ref{selfs}) can be written as
\beq
F(t,t') = \left[ \frac{\delta^2 Z}{\delta \Re h(t) \delta \Re h(t')} + \frac{\delta^2 Z}{\delta \Im h(t) \delta \Im h(t')} - \frac{\delta Z}{\delta \Re h(t)} \frac{\delta Z}{\delta \Re h(t')} - \frac{\delta Z}{\delta \Im h(t)} \frac{\delta Z}{\delta \Im h(t')} \right]_{h=0}
\eeq
using the symmetry under permutations of the $\varphi_i$.
Defining formally the operators
\beq
\label{derivatestrane}
\frac{\delta}{\delta h(t)}=\frac{\delta}{\delta \Re h(t)} + I \frac{\delta}{\delta \Im h(t)} \hspace{1cm} \frac{\delta}{\delta h^*(t)} = \left(\frac{\delta}{\delta h(t)}\right)^*
\eeq
one can see that
\beq
F(t,t')= \left[ \frac{\delta^2 Z}{\delta h(t) \delta h^*(t')} -  \frac{\delta Z}{\delta h(t)} \frac{\delta Z}{\delta h^*(t')} \right]_{h=0} 
\eeq
using $\langle \sin(\varphi_i(t))\cos(\varphi_i(t')) \rangle \equiv 0$ and $\langle \sin(\varphi_i(t)) \rangle \equiv 0$ because of the symmetry $\varphi \rightarrow -\varphi$, as we have already discussed before Eq.~(\ref{selfs}).
We can generalize this relation defining the supercorrelator
\beq
\label{selfcorrdef}
\begin{split}
F(a,b)&= \frac{1}{N} \sum_i \left[ \langle e^{I(\phi_i(a)-\phi_i(b))} \rangle - \langle e^{I \phi_i(a)} \rangle \langle e^{-I \phi_i(b) } \rangle \right] \\
& = \left[ \frac{\delta^2 Z}{\delta h(a) \delta h^*(b)} - \frac{\delta Z}{\delta h(a)} \frac{\delta Z}{\delta h^*(b)} \right]_{h=0} 
\end{split}
\eeq

\subsection{The saddle-point equations}

\noindent
Substituting Eq.~(\ref{hamiltonian}) in Eq.~(\ref{genfunct}) and introducing the supervariable
\beq
z(a)=\frac{1}{N} \sum_i e^{I\phi_i(a)}
\eeq
we get
\beq
Z[0] = 1 = \int {\cal D}\phi_i {\cal D}z \  \delta \left( N z(a)- \sum_i e^{i\phi_i(a)} \right)  \ \exp \left[ \ \sum_i T(\phi_i)  - N \Delta \int da \ (1-\Re z(a)^k) \right]
\eeq
where $T(\phi)=\frac{1}{2} \int da \ \phi \Delta^{(2)} \phi$.
Using the integral representation of the $\delta$-function
\beq
\delta(z)=\int {\cal D}\hat{z} \exp \left[ \int da \ \Re(z(a) \ i \hat{z}(a)) \right]
\eeq
we get, rotating as usual the integration path in the $\hat{z}$ plane,
\beq
\label{saddlefunct}
\begin{split}
Z[0] &= \int {\cal D}\phi_i {\cal D}z {\cal D}\hat{z} \ \exp \left[ \ N \Re \int da \ \hat{z}(a) \ z(a) - \sum_i \Re \int da \ \hat{z}(a) \ e^{I\phi_i(a)}+ \sum_i T(\phi_i)  - N \Delta \Re \int da \ (1-z(a)^k) \right] \\
&=  \int {\cal D}z {\cal D}\hat{z} \ \exp N \left[ \ \Re \int da \ \hat{z}(a) \ z(a)  - \Delta \Re \int da \ (1-z(a)^k) + \log {\cal Z}[\hat{z}] \right] \\
&=  \int {\cal D}z {\cal D}\hat{z} \ \exp N L(z,\hat{z})
\end{split}
\eeq
where we defined
\beq
{\cal Z}[\hat{z}]=\int {\cal D}\phi \ \exp \left[ T(\phi) - \Re \int da \ \hat{z}(a) \ e^{I\phi(a)} \right]
\eeq
By comparison with Eq.~(\ref{genfunct}) we see that $Z[\hat{z}]$ is the generating functional for the dynamics of a single degree of freedom with energy
\beq
\label{heffsuperfield}
{\cal H}(\phi,\hat{z})=\Re \hat{z}(a) e^{I\phi}
\eeq
We can evaluate the integral in Eq.~ (\ref{saddlefunct}) by a saddle point, as usual in mean field models; we get, using again the formal operators (\ref{derivatestrane})
\beq
\label{eqsaddle}
\begin{split}
\frac{\delta L}{\delta z^*(a)}= \hat{z}(a) + \Delta k z(a)^{k-1} = 0 \ \ &\Longrightarrow \ \ \hat{\zeta}(a)=- \Delta k \zeta(a)^{k-1} \\
\frac{\delta L}{\delta \hat{z}^*(a)}= z(a) + \frac{\delta}{\delta \hat{z}^*(a)} \log {\cal Z}[\hat{z}] = 0 \ \ &\Longrightarrow \ \ \zeta(a)= \langle e^{I\phi(a)} \rangle_{{\cal H}(\hat{\zeta})}
\end{split}
\eeq
where we defined $\zeta$ and $\hat{\zeta}$ as the saddle point values of $z$ and $\hat{z}$ respectively.
So, in the thermodynamic limit the dinamics of the system is equivalent to the one of a single degree of freedom with hamiltonian
\beq
\label{heffsusy}
\begin{cases}
{\cal H}(\phi,a)=-\Delta \Re [ k \zeta(a)^{k-1} e^{I\phi} ] \\
\zeta(a)=\langle e^{I\phi(a)} \rangle
\end{cases}
\eeq
Setting $\theta=\bar{\theta}=0$, we get the effective dynamical system for the variable $\varphi$:
\beq
\label{heffcorretta}
\begin{cases}
{\cal H}(\varphi,t)=-\Delta \Re [ k \zeta(t)^{k-1} e^{I\varphi} ] \\
\zeta(t)=\langle e^{I\varphi(t)} \rangle
\end{cases}
\eeq

\subsection{Self correlation functions}

\noindent
We want now to show that it is possible to use the effective hamiltonian, Eq.~(\ref{heffcorretta}), to calculate the self correlation function given by Eq.~(\ref{selfcorrdef}).
In fact we obtain in the same way as we obtained Eq.~(\ref{saddlefunct})
\beq
\begin{split}
Z[h]
&=  \int {\cal D}z {\cal D}\hat{z} \ \exp \Bigl[ \ N \ \Re \int da \ \hat{z}(a) \ z(a)  - N \Delta \Re \int da \ (1-z(a)^k)  \\
&+ (N-1) \log {\cal Z}[\hat{z}] + \log {\cal Z}[\hat{z} + h] \Bigr] \\
&= \int {\cal D}z {\cal D}\hat{z} \ \exp \Bigl[ N L(z,\hat{z}) + \log {\cal Z}[\hat{z} + h] - \log {\cal Z}[\hat{z}] \Bigr]
\end{split}
\eeq
so that
\beq
F(a,b)= \int {\cal D}z {\cal D}\hat{z} \  \frac{\delta^2 \log {\cal Z}[\hat{z}]}{\delta \hat{z}(a) \delta \hat{z}^*(b)}   \exp \left [ N L(z,\hat{z}) \right]
\eeq
In the termodynamic limit we know from the saddle-point equations that the integral is dominated by $z=\zeta$, $\hat{z}=\hat{\zeta}$, and that
\beq
\exp N L(\zeta,\hat{\zeta}) \sim {\cal Z}[0] \sim 1
\eeq
so that
\beq
\label{selfuguali}
F(a,b)= \left[ \frac{\delta^2 \log {\cal Z}[\hat{z}]}{\delta \hat{z}(a) \delta \hat{z}^*(b)} \right]_{\hat{z}=\hat{\zeta}} = \langle e^{I ( \phi(a) - \phi(b) )} \rangle_{{\cal H}(\hat{\zeta})} - \langle e^{I \phi(a)} \rangle_{{\cal H}(\hat{\zeta})}  \langle e^{-I \phi(b)} \rangle_{{\cal H}(\hat{\zeta})} \equiv {\cal F}(a,b)
\eeq
The self correlation function is then equal to the one calculated for a single degree of freedom using the effective hamiltonian (\ref{heffcorretta}). The same arguments is extended to n-times correlations by differentiating n-times.

%%%%%%%%%%%%%%%%%%%%%%%%%%%%%%%%%%%%%%%%%%%%%%%%%%%%%%%%%%%%%%%%%%%%%%%%%%%%%%%%%%%

\section{Collective dynamics}
\label{app_coll}

\noindent
In this appendix we will derive a general relation between the self correlation functions and the collective ones. The latter vanish in the thermodynamic limit, and are related to the Gaussian corrections around the mean-field saddle point that we studied in appendix \ref{app_self}.

\subsection{Gaussian corrections to mean field}

\noindent
We want now to calculate the correlation function of the ``magnetization'' (multiplied by $N$ in order to have a well defined thermodynamic limit)
\beq
G(a,b)=N [\langle z(a) z^*(b) \rangle - \langle z(a) \rangle \langle z^*(b) \rangle]
\eeq
by expanding $L$ at second order around the saddle point; to do this, it is not possible to use the operators (\ref{derivatestrane}). We have to separate the real and imaginary part of $z$ and $\hat{z}$. We get, calling $\Delta z_\sigma = [\Re(z-\zeta),\Im(z-\zeta),\Re(\hat{z} - \hat{\zeta}),\Im(\hat{z} - \hat{\zeta})]$ with $\sigma=1,2,3,4$:
\beq
L(z,\hat{z})=L(\zeta,\hat{\zeta}) + \frac{1}{2} \ \left[ \sum_{\sigma,\sigma'} \ \int \  da \  db \ \Delta z_\sigma(a) \  \frac{\delta^2 L}{\delta z_\sigma(a) \delta z_{\sigma'}(b)} \ \Delta z_{\sigma'}(b) \right]
\eeq
Then, at second order around the saddle point,
\beq
P(\Delta z) \propto \exp \left[ - \frac{N}{2} \ \sum_{\sigma,\sigma'} \ \int \  da \  db \ \Delta z_\sigma(a) \ J_{\sigma \sigma'}(a,b) \ \Delta z_{\sigma'}(b) \right]
\eeq
where the matrix $J(a,b)$ is given by
\beq
J(a,b)= - 
\begin{bmatrix}
\Re w(a) \ \delta(a-b) & -\Im w(a) \ \delta(a-b) & \delta(a-b) & 0   \\
 -\Im w(a) \ \delta(a-b) & -\Re w(a) \ \delta(a-b) & 0 & - \delta(a-b)  \\
\delta(a-b) & 0 &   {\cal F}_c(a,b)   & 0  \\
0    &  -\delta(a-b)  &   0  & {\cal F}_s(a,b)
\end{bmatrix}
\eeq
where
\beq
\begin{split}
{\cal F}_c(a,b)=&\langle \cos(\phi(a)) \cos(\phi(b)) \rangle_{{\cal H}(\hat{\zeta})}  - \langle \cos(\phi(a)) \rangle_{{\cal H}(\hat{\zeta})}  \langle \cos(\phi(b)) \rangle_{{\cal H}(\hat{\zeta})} \\
&\langle \cos(\phi(a)) \sin(\phi(b)) \rangle_{{\cal H}(\hat{\zeta})} = 0 \\
{\cal F}_s(a,b)=&\langle \sin(\phi(a)) \sin(\phi(b)) \rangle_{{\cal H}(\hat{\zeta})} - \langle \sin(\phi(a)) \rangle_{{\cal H}(\hat{\zeta})}  \langle \sin(\phi(b)) \rangle_{{\cal H}(\hat{\zeta})} \\
w(a)=&k(k-1) \zeta(a)^{k-2} 
\end{split}
\eeq
are calculated on the effective hamiltonian (\ref{heffsusy}).
We have also
\beq
J_{\sigma \sigma'}(a,b)=J_{\sigma' \sigma}(b,a)
\eeq
Then defining
\beq
G_{\sigma \sigma'} = N \langle \Delta z_\sigma(a) \Delta z_{\sigma'}(b) \rangle
\eeq
one has
\beq
\label{inverti}
\sum_{\sigma'} \int db J_{\sigma \sigma'}(a,b) G_{\sigma' \sigma''}(b,c) = \delta_{\sigma \sigma''} \delta(a-c)
\eeq
We are interested in calculating
\beq
\begin{split}
G(a,b) &=N [\langle z(a) z^*(b) \rangle - \langle z(a) \rangle \langle z^*(b) \rangle] = N\langle \Delta z_1(a) \Delta z_1(b) \rangle + N\langle \Delta z_2(a) \Delta z_2(b) \rangle \\ &= G_{11}(a,b) + G_{22}(a,b)
\end{split}
\eeq
By writing explicitly some of the Eq.s~(\ref{inverti}) and making some substitutions one gets
\beq
\begin{split}
\label{esplicite}
&G_{11}(a,c)= {\cal F}_c(a,c) + \int db \ {\cal F}_c(a,b) \  \left[ \Re w(b) G_{11}(b,c) - \Im w(b) G_{21}(b,c) \right] \\
&G_{12}(a,c)= \int db \ {\cal F}_c(a,b) \ \left[\Re w(b) G_{12}(b,c) - \Im w(b) G_{22}(b,c) \right] \\
&G_{21}(a,c)= - \int db \ {\cal F}_s(a,b) \ \left[\Im w(b) G_{11}(b,c) + \Re w(b) G_{21}(b,c) \right] \\
&G_{22}(a,c)=  {\cal F}_s(a,c) - \int db \  {\cal F}_s(a,b) \ \left[ \Im w(b) G_{12}(b,c) + \Re w(b) G_{22}(b,c) \right]
\end{split}
\eeq
These equations give the collective correlation functions in term of the self correlations evaluated on the effective hamiltonian; but using Eq.~(\ref{selfuguali}) we can replace the effective self correlations with the original ones, and obtain a relation between self and collective correlations of the original system.

\subsection{Solution in the equilibrium case}

\noindent
Probability conservation and causality imply that \cite{cugliandolo2,cugliandolo1}
\beq
\label{sempliceforma}
\begin{split}
G_{\sigma \sigma'}(a,b)&=C_{\sigma \sigma'}(t,t') + (\bar{\theta}'-\bar{\theta})\bigl( \theta' R_{\sigma \sigma'}(t,t') +  \theta \bar{R}_{\sigma \sigma'}(t,t') \bigr) \\
{\cal F}_{c,s}(a,b)&= {\cal C}_{c,s}(t,t') + (\bar{\theta}'-\bar{\theta})\bigl( \theta' {\cal R}_{c,s}(t,t') +  \theta \bar{\cal R}_{c,s}(t,t') \bigr)
\end{split}
\eeq
where
\beq
\bar{R}(t,t')=R(t',t)
\eeq
In the $t \rightarrow \infty$ limit the system reaches equilibrium, so that the correlation function depend only on the time difference, the response function are related to the correlations by the fluctuation-dissipation theorem, and $w$ does not depend on time and is equal to the equilibrium magnetization. From the solution of the thermodynamics we know that the equilibrium magnetization does not depend on $\theta$ and $\bar{\theta}$, so that
\beq
\lim_{t \rightarrow \infty} w(a) = w
\eeq
and
\beq
\lim_{\substack{t,t' \rightarrow \infty \\ t-t'=\tau}} G_{\sigma \sigma'}(a,b)=
C_{\sigma \sigma'}(t-t') + (\bar{\theta}'-\bar{\theta})\bigl( \theta' R_{\sigma \sigma'}(t-t') +  \theta R_{\sigma \sigma'}(t'-t) \bigr)
\eeq
If we suppose that the correlations decay fast (exponentially) for $\tau \rightarrow \infty$, the values of $t_b$ in the integrals in Eq.~(\ref{esplicite}) must stay close to $t_a$, $t_c$. Then in the limit in which $t_a$, $t_c$ go to infinity also $t_b$ must go to infinity, and $w$ becomes a constant also with respect to the fermionic variables. Then we can rewrite Eq.~(\ref{esplicite}) 
\beq
\begin{split}
\label{espliciteinfinito}
&G_{11}= {\cal F}_c + \Re w \ {\cal F}_c \varotimes G_{11} - \Im w \ {\cal F}_c \varotimes G_{21}  \\
&G_{12}= \Re w \ {\cal F}_c \varotimes G_{12} - \Im w \ {\cal F}_c \varotimes G_{22}  \\
&G_{21}= - \Im w \ {\cal F}_s \varotimes G_{11} - \Re w \ {\cal F}_s \varotimes G_{21}  \\
&G_{22}=  {\cal F}_s - \Im w \ {\cal F}_s \varotimes G_{12} - \Re w \ {\cal F}_s \varotimes G_{22} 
\end{split}
\eeq
where
\beq
(G_1 \varotimes G_2)(a,c)= \int db \ G_1(a,b) \ G_2(b,c)
\eeq
If $G_1$ and $G_2$ are of the form (\ref{sempliceforma}), $G_1 \varotimes G_2$ has the same form \cite{cugliandolo2,cugliandolo1}:
\beq
(G_1 \varotimes G_2)(a,c)= (C_1 \otimes \bar{R}_2 + R_1 \otimes C_2)(t,t') + (\bar{\theta}'-\bar{\theta}) \bigl(
 \theta' (R_1 \otimes R_2)(t,t') + \theta (\bar{R}_2 \otimes \bar{R}_1)(t,t') \bigl)
\eeq
where $\otimes$ is the time convolution product.
Equating (for example) the $\theta' \bar{\theta}'$ component of Eq.~(\ref{espliciteinfinito}) and moving to the frequency domain we finally get
\beq
\begin{split}
&R_{11}(\omega)= {\cal R}_c(\omega) + \Re w \ {\cal R}_c(\omega) R_{11}(\omega) - \Im w \ {\cal R}_c(\omega) R_{21}(\omega) \\
&R_{12}(\omega)= \Re w \ {\cal R}_c(\omega) R_{12}(\omega) - \Im w \ {\cal R}_c(\omega) R_{22}(\omega)  \\
&R_{21}(\omega)= - \Im w \ {\cal R}_s(\omega) R_{11}(\omega) - \Re w \ {\cal R}_s(\omega) R_{21}(\omega)  \\
&R_{22}(\omega)=  {\cal R}_s(\omega) - \Im w \ {\cal R}_s(\omega) R_{12}(\omega) - \Re w \ {\cal R}_s(\omega) R_{22}(\omega) 
\end{split}
\eeq
and
\beq
\label{relgencollself}
\begin{split}
&R_{11}(\omega)=\frac{ {\cal R}_c(\omega)}{1 - \Re w \ {\cal R}_c(\omega) + \frac{(\Im w)^2 {\cal R}_c(\omega) {\cal R}_s(\omega)}{1+ \Re w \ {\cal R}_s(\omega)}} \\
&R_{22}(\omega)=\frac{ {\cal R}_s(\omega)}{1 + \Re w \ {\cal R}_s(\omega) - \frac{(\Im w)^2 {\cal R}_c(\omega) {\cal R}_s(\omega)}{1- \Re w \ {\cal R}_c(\omega)}} 
\end{split}
\eeq
Using the fluctuation-dissipation theorem, one can transform this relation in a relation between the correlation functions.

\subsection{Other simplifications}

\noindent In our model one can always choose the magnetization such that $\Im w=0$. In this case one has
\beq
\begin{split}
&R_{11}(\omega)=\frac{ {\cal R}_c(\omega)}{1 - w \ {\cal R}_c(\omega)} \\
&R_{22}(\omega)=\frac{ {\cal R}_s(\omega)}{1 + w \ {\cal R}_s(\omega)}
\end{split}
\eeq
Moreover, we can assume that the self correlation functions are exponentials, so that
\beq
{\cal R}_{c,s}(\omega)=\frac{\beta A_{c,s}}{1-i \omega \tau_{c,s}}
\eeq
We get easily
\beq
R_{11}(\omega) = \frac{ \beta Z_c A_c}{1-i \omega Z_c \tau_c} \hspace{1cm} Z_c=(1-\beta w A_c)^{-1}
\eeq
The same relation holds for $R_{22}$ with the substitution $w \rightarrow -w$. From this relations, using the fluctuation-dissipation theorem and moving back to the time domain, one gets Eq.s~(\ref{GGG}).

%%%%%%%%%%%%%%%%%%%%%%%%%%%%%%%%%%%%%%%%%%%%%%%%%%%%%%%%%%%%%%%%%%%%%%%%%%%%%%%%%%%%%%%%%%%%%%%%%%%%%%%%%%%%%%%%%%%%%%%%%%%%%%%%%%%%%%%%%%%%%%%%%%%%%%%%%%%%%%

\section{Effective dynamical system for the minimization of $W$}
\label{app_W}

\noindent
The calculation of the effective potential ${\cal W}$ used in Eq.~(\ref{Wsingola}) is carried on the same lines of the one presented in section 2 of appendix \ref{app_self}. We start from the generating functional
\beq
Z[0]=\int {\cal D}\phi_i \ \exp \left[ \ \frac{1}{2} \sum_i \int \  da \ \phi_i(a) \ \Delta^{(2)} \ \phi_i(a) - \int da \ W(\phi)  \ \right]
\eeq
where $W(\phi)$ is given by Eq.~(\ref{W}). Now we introduce $\delta$ functions for the variables $z(a)$ and $z_2(a)$, and we get
\beq
Z[0]=\int {\cal D}z {\cal D}\hat{z} {\cal D}z_2 {\cal D}\hat{z}_2 \ \exp N \left[\ \Re \int da ( z(a) \hat{z}(a) +  z_2(a) \hat{z}_2(a))  - \int da \ \frac{W(z,z_2)}{N}  + \log{\cal Z}(\hat{z},\hat{z}_2) \ \right]
\eeq
where now 
\beq
{\cal Z}(\hat{z},\hat{z}_2) = \int {\cal D}\phi \ \exp \left[ \ T(\phi) - \Re \int da \ \left( \ \hat{z}(a) \ e^{I\phi(a)} + \hat{z}_2(a) \ e^{2 I \phi(a)} \ \right) \ \right]
\eeq
We find then that
\beq
{\cal W}(\phi,a)=\Re \left[ \hat{z}(a) \ e^{I\phi} + \hat{z}_2(a) \ e^{2 I \phi} \right]
\eeq
The saddle point equation are
\beq
\begin{split}
&\frac{\delta L}{\delta z^*(a)}= \hat{z}(a) - \frac{1}{N} \frac{\delta W}{\delta z^*(a)} = 0 \ \ \Longrightarrow \ \ \hat{\zeta}(a)= \Delta^2 k^2 (k-1) \ [ \ \zeta^*(a) \ (\zeta(a) \zeta^*(a))^{k-2} - \zeta_2(a) \ \zeta(a)^{2k-3} \ ]  \\
&\frac{\delta L}{\delta \hat{z}^*(a)}= z(a) + \frac{\delta}{\delta \hat{z}^*(a)} \log {\cal Z}[\hat{z},\hat{z}_2] = 0 \ \ \Longrightarrow \ \ \zeta(a)= \langle e^{I\phi(a)} \rangle_{{\cal W}(\hat{\zeta},\hat{\zeta}_2)} \\
&\frac{\delta L}{\delta z_2^*(a)}= \hat{z_2}(a) - \frac{1}{N} \frac{\delta W}{\delta z_2^*(a)} = 0 \ \ \Longrightarrow \ \ \hat{\zeta}_2(a)= - \frac{1}{2} \Delta^2 k^2  \zeta(a)^{2k-2}  \\
&\frac{\delta L}{\delta \hat{z}_2^*(a)}= z_2(a) + \frac{\delta}{\delta \hat{z}_2^*(a)} \log {\cal Z}[\hat{z},\hat{z}_2] = 0 \ \ \Longrightarrow \ \ \zeta_2(a)= \langle e^{2I\phi(a)} \rangle_{{\cal W}(\hat{\zeta},\hat{\zeta}_2)}
\end{split}
\eeq
and finally we obtain
\beq
\begin{cases}
{\cal W}(\phi,a) = \Delta^2 k^2 (k-1) \ \Re \{ \ [ \ \zeta^*(a) \ (\zeta(a) \zeta^*(a))^{k-2} - \zeta_2(a) \ \zeta(a)^{2k-3} \ ] \ e^{I \phi} \ \} - \frac{1}{2} \Delta^2 k^2 \ \Re \{ \ \zeta(a)^{2k-2} \ e^{2 I \phi} \ \} \\
 \zeta(a)= \langle e^{I\phi(a)} \rangle \\
 \zeta_2(a)= \langle e^{2I\phi(a)} \rangle
\end{cases}
\eeq
that after setting $\theta=\bar{\theta}=0$ reduces to Eq.~(\ref{Wsingola}).

\section{Closest saddles to equilibrium configurations}
\label{app_distance}

\noindent
In this section we will derive the result presented in section \ref{Wvsdistance}. We have to compute the quantity
\beq
\sigma(T;e_s,q)=\frac{1}{N} \int d\varphi_i \frac{e^{-\beta H(\varphi)}}{Z(T)} \log \int d \psi_i \ \delta(H(\psi)-Ne_s) \ \delta(\partial_i H(\psi)) \ \det \text{H}(\psi) \ \delta \left( q - q(\varphi,\psi) \right) \ ,
\eeq
where $q(\varphi,\psi)=N^{-1} \sum_i \cos( \varphi_i - \psi_i)$.
To do that, we need to prove a general relation. Suppose we want to calculate at the saddle point a quantity $Q$ of the form
\beq
Q=\frac{1}{N} \int d\varphi_i \frac{e^{-\beta H(\varphi)}}{Z(T)} \log A(\varphi) = \lim_{n \rightarrow 0} \frac{1}{Nn}\left[ \int d\varphi_i \frac{e^{-\beta H(\varphi)}}{Z(T)} A^n(\varphi) - 1 \right] = \lim_{n \rightarrow 0} \frac{1}{Nn} \log \int d\varphi_i \frac{e^{-\beta H(\varphi)}}{Z(T)} A^n(\varphi) \ ,
\eeq
where we used the relations $\log x = \lim_{n \rightarrow 0} \frac{x^n -1}{n}$ and $\lim_{n \rightarrow 0} (f(n)-1) = \lim_{n \rightarrow 0} \log f(n)$ if $f(n) \rightarrow_{n \rightarrow 0} 1$.
Suppose also that the energy depends only on some collective parameter as in mean field models: $H(\varphi)=Ne(z(\varphi))$, where $Nz(\varphi)=\sum_i z(\varphi_i)$ (in our model, $z(\varphi_i)=\exp(I\varphi_i)$). Then we have
\beq
\begin{split}
Q=&\lim_{n \rightarrow 0} \frac{1}{Nn} \log \int dz \frac{e^{-\beta N e(z)}}{Z(T)} \int d\varphi_i \ \delta(z-z(\varphi)) \ A^n(\varphi) \\
=&\lim_{n \rightarrow 0} \frac{1}{Nn} \log \int dz d\hat{z} \frac{e^{-\beta N e(z)}}{Z(T)} \int d\varphi_i \ e^{ i\hat{z} (N z-\sum_i z(\varphi_i))} \ A^n(\varphi) \\
=& \lim_{n \rightarrow 0} \frac{1}{Nn} \log \frac{1}{Z(T)} \int dz d\hat{z} \ e^{-\beta N (e(z) -T s(n;z,i\hat{z}))} \ ,
\end{split}
\eeq
where we defined
\beq
s(n;z,i\hat{z}) = z \ i\hat{z} + \frac{1}{N} \log  \int d\varphi_i \ e^{- i\hat{z}  \sum_i z(\varphi_i)} \ A^n(\varphi) \ .
\eeq
Clearly $s(0;z,i\hat{z})$ is the entropic contribution to the free energy as a function of $z$, $\hat{z}$ that we obtain in the calculation of the partition function $Z(T)$, so that
\beq
f(T)=-\frac{1}{\beta N} \log Z(T) = \min_{z,\hat{z}} [ e(z) - T s(0;z,i\hat{z}) ] = e(\zeta) - T s(0;\zeta,\hat{\zeta})=f(0;\zeta,\hat{\zeta}) \ ,
\eeq
where $(\zeta(T),\hat{\zeta}(T))$ is the (T-dependent) thermodynamic minimum of the free energy (note that at the saddle point $i\hat{z}=\hat{\zeta}$).
Then we have
\beq
Q= \lim_{n \rightarrow 0} \frac{1}{Nn} \log \int dz \ e^{-\beta N [ f(n;z,i\hat{z}) - f(0;\zeta,\hat{\zeta}) ]} \ .
\eeq
We can now expand $z = \zeta + n \zeta^{(1)} + o(n^2)$, $i\hat{z} = \hat{\zeta} + n \hat{\zeta}^{(1)} + o(n^2)$ and
\beq
f(n;z,i\hat{z}) - f(0;\zeta,\hat{\zeta}) = \frac{\partial f}{\partial z}(0;\zeta,\hat{\zeta}) \ n \zeta^{(1)} +  \frac{\partial f}{\partial i\hat{z}}(0;\zeta,\hat{\zeta}) \ n \hat{\zeta}^{(1)} + \frac{\partial f}{\partial n}(0;\zeta,\hat{\zeta}) \ n + o(n^2) = \frac{\partial f}{\partial n}(0;\zeta,\hat{\zeta}) \ n + o(n^2)   \ ,
\eeq
because by definition of ($\zeta$,$\hat{\zeta}$) we have $\frac{\partial f}{\partial z}(0;\zeta,\hat{\zeta})=0$, $\frac{\partial f}{\partial i\hat{z}}(0;\zeta,\hat{\zeta})=0$. We get then the final result:
\beq
Q= -\beta \frac{\partial f}{\partial n}(0;\zeta,\hat{\zeta}) =  \frac{\partial s}{\partial n}(0;\zeta,\hat{\zeta}) \ .
\eeq
We have then to calculate (neglecting the term $\zeta \hat{\zeta}$ that vanish on taking the derivative with respect to $n$):
\beq
s(n;\hat{\zeta},e_s,q) = \frac{1}{N} \log  \int d\varphi_i \ e^{-\sum_i \hat{\zeta} \cos \varphi_i} \ \prod_{a=1}^n \int d \psi^a_i \ \delta(H(\psi^a)-Ne_s) \ \delta(\partial_i H(\psi^a)) \ \det \text{H}(\psi^a) \ \delta \left( q - q(\varphi,\psi^a) \right) \ .
\eeq
where from the thermodynamic calculation $\hat{\zeta}(T)=-\beta k \zeta^{k-1}$ and $\zeta$ is given by  given by Eq.~(\ref{zeta}).
Using a representation analogous to Eq.~(\ref{chi_E_part}) we get
\beq
\begin{split}
s(n;\hat{\zeta},e_s,q) = &\frac{1}{N} \log  \int d\varphi_i \  e^{- \sum_i \hat{\zeta} \cos \varphi_i} \\
& \prod_{a=1}^n \int \frac{d \beta_a}{2\pi} e^{N \beta_a e_s} \int {\cal D}\Psi^a_i \  \exp \left[ \int d\bar{\theta} d\theta (1-\beta_a \theta \bar{\theta}) H(\Psi^a) \right] \ \delta \left( N q - \sum_i \cos(\varphi_i - \psi_i^a) \right) \ .
\end{split}
\eeq
We will now: {\it i)} substitute the expression $H(\Psi^a)=N(1-\Re y_a^k)$, using $y$ instead of $z$ to avoid confusion with the thermodynamic variable $\zeta$; {\it ii)} insert some $\delta$-functions for $y_a$ and the corresponding integral representation with a multiplier $\hat{y}_a$; {\it iii)} neglect all the product and sum signs related to the index $a$; {\it iv)} use the integral representation for the $\delta$-function of $q$ with a multiplier $\lambda_a$. Then we get an expression that has to be maximized with respect to all the parameters to get the saddle point value of $s(n;\hat{\zeta},e_s,q)$:
\beq
\begin{split}
&s(n;\hat{\zeta},e_s,q)=\max_{\text{all par}} \left[ \sum_a \beta_a e_s + \sum_a \int d\bar{\theta} d \theta [ (1- \beta_a \theta \bar{\theta}) (1 - \Re y_a^k) + \Re y_a \hat{y}_a ] + \sum_a \lambda_a q + \log {\cal S}(\hat{\zeta},\hat{y}_a,\lambda_a) \right] \\
&{\cal S}(\hat{\zeta},\hat{y}_a,\lambda_a) = \int d\varphi \ {\cal D}\Psi^a \exp \left[ -\hat{\zeta} \cos \varphi - \sum_a \int d\bar{\theta} d \theta \ \Re \hat{y}_a e^{I \Psi^a} - \sum_a \lambda_a \cos(\varphi-\psi^a) \right] 
\end{split}
\eeq
As usual, we will assume that: {\it i)} there is symmetry between the replicas ($y_a=y$, etc.); {\it ii)} $y$ and $\hat{y}$ are real; {\it iii)} all the fermionic components are 0. Then we get
\beq
\begin{split}
&s(n;\hat{\zeta},e_s,q)=\max_{\text{all par}} \left[ n \Big( \beta (e_s - 1 + y_0^k) - k y_0^{k-1} y_3  + \hat{y}_0 y_3 + \hat{y}_3 y_0 + \lambda q \Big) + \log {\cal S}(\hat{\zeta},\hat{y},\lambda) \right] \\
&{\cal S}(\hat{\zeta},\hat{y},\lambda) = \int d\varphi \ e^{ -\hat{\zeta} \cos \varphi} \left[ \int {\cal D}\Psi \exp \Big( -  \int d\bar{\theta} d \theta \ (\hat{y}_0 + \hat{y}_3 \theta \bar{\theta}) \cos \Psi -  \lambda \cos(\varphi-\psi) \Big) \right]^n 
\end{split}
\eeq
Now we have to take the derivative of $s$ with respect to $n$ at $n=0$. By direct computation
\beq
\begin{split}
\sigma(\hat{\zeta};e_s,q)=\max_{\text{all par}} \frac{\partial s}{\partial n}(0;\hat{\zeta},e_s,q)&=\max_{\text{all par}} \Big[  \beta (e_s - 1 + y_0^k) - y_3 (k y_0^{k-1} - \hat{y}_0) + \hat{y}_3 y_0 + \lambda q \\
&+ \int d \varphi \frac{e^{-\hat{\zeta} \cos \varphi}}{2 \pi I_0(\hat{\zeta})} \log  \int {\cal D}\Psi \exp \Big( -  \int d\bar{\theta} d \theta \ (\hat{y}_0 + \hat{y}_3 \theta \bar{\theta}) \cos \Psi -  \lambda \cos(\varphi-\psi) \Big) \Big]
\end{split}
\eeq
The interpretation of this expression is straightforward by comparison with Eq.~(\ref{sigma_top}): in fact if we put $\lambda=0$ we get exactly Eq.~(\ref{sigma_top}). This correspond to integrating $\sigma$ over $q$, so the dependence on the reference configuration (and hence on the temperature) disappears and we get the number of saddles of energy $e_s$. When $\lambda$ is different from 0 the last term of the previous expression represents the single-particle version of $\sigma$. Now we can proceed exactly in the same way as we proceeded after Eq.~(\ref{sigma_top}): we take the derivatives with respect to $\beta$ and $y_3$. This fixes $y_0=(1-e_s)^{1/k}$ and $\hat{y}_0=k y_0^{k-1}$ and equals to zero the first two terms of $\sigma$. As we are looking for saddles of energy $e_s<1$, we have then $\hat{y}_0 > 0$, and then the dependence on $\hat{y}_0$ in the last term disappears. We get 
\beq
\sigma(\hat{\zeta};e_s,q)=\max_{\lambda,\hat{y}_3} \Big[ \hat{y}_3 y_0 + \lambda q
+ \int d \varphi \ {\cal P}(\varphi) \ \log (-2 \sinh (\hat{y}_3 + \lambda \cos \varphi))  \Big]
\eeq
where ${\cal P}(\varphi)$ is given by Eq.~(\ref{gibbs}).
From the equation $y_0=(1-e_s)^{1/k}$ we see that $y_0$ is the average of $\cos \varphi$ on the saddles; then we will change the notation calling $y_0=\zeta_s$. Taking the derivatives of $\sigma$ with respect to $\lambda$ and $\hat{y}_3$ we get
\beq
\begin{split}
\zeta_s=&-\int d \varphi \ {\cal P}(\varphi) \ [ \tanh u(\varphi) ]^{-1} \\
q=&-\int d \varphi \ {\cal P}(\varphi) \ \cos \varphi \ [ \tanh u(\varphi) ]^{-1}
\end{split}
\eeq
where $u(\varphi)=\hat{y}_3 + \lambda \cos \varphi$.
Now if we want that $\zeta_s \in [0,1]$, $u(\varphi)$ must have an imaginary part; but if we want $\zeta_s$ to be real, this imaginary part must be constant and equal to $\pi/2$.  We will then assume that $\hat{y}_3 = y + i\frac{\pi}{2}$; note that this is the correct solution for $\lambda=0$ (see Eq.~(\ref{hatz3eul})). We obtain easily the parametric relation for $\sigma(T;\zeta_s,q)$:
\beq
\begin{split}
\zeta_s(y,\lambda)=&-\int d \varphi \ {\cal P}(\varphi) \ f(t(\varphi)) \\
q(y,\lambda)=&-\int d \varphi \ {\cal P}(\varphi) \ \cos \varphi \ f(t(\varphi)) \\
\sigma(y,\lambda)=&\int d \varphi \ {\cal P}(\varphi) \Big[ \log 2 \cosh t(\varphi) - t(\varphi) f(t(\varphi)) \Big] - i\pi \frac{1-\zeta_s(y,\lambda)}{2}
\end{split}
\eeq
where $t(\varphi)=y+\lambda \cos \varphi$ and $f(t)=\frac{\cosh t -1}{\sinh t}$. We see that the imaginary part of $\sigma$ is, as expected, equal to $i \pi n(e_s)$, remembering the relation between energy and order of each stationary point. In the following we will neglect the imaginary part of $\sigma$. Now we have to maximize $q$ on the curve $\sigma=0$ (see section \ref{Wvsdistance}). To do that, we start with a simple argument: as $\lambda$ is the field conjugated to $q$ (the relation between $\lambda$ and $q$ is the same as the relation between magnetization and magnetic field in a ferromagnet) we expect that the maximum overlap will be obtained in the $|\lambda| \rightarrow \infty$ limit. In fact, for $\lambda \rightarrow \pm \infty$, we have $t(\varphi) \rightarrow \pm \text{sgn}(\cos \varphi) \ \infty$ and $f(t(\varphi)) \rightarrow \pm \text{sgn}(\cos \varphi)$, so that
\beq
\lim_{\lambda \rightarrow \pm \infty} q(y,\lambda) = \mp \int d \varphi \ {\cal P}(\varphi) \ | \cos \varphi | \ .
\eeq
As we want $q$ to be positive, we have to choose $\lambda \rightarrow -\infty$. We have then
\beq
\label{limit_lambda}
\begin{split}
&\lim_{\lambda \rightarrow - \infty} q(y,\lambda)  = \int d \varphi \ {\cal P}(\varphi) \ | \cos \varphi | = \bar{q} \ , \\
&\lim_{\lambda \rightarrow - \infty} \zeta(y,\lambda)  = \int d \varphi \ {\cal P}(\varphi) \  \text{sgn}(\cos \varphi) = \bar{\zeta}_s \ , \\
&\lim_{\lambda \rightarrow - \infty} \sigma(y,\lambda) = 0 \ .
\end{split}
\eeq
Then we have a consistency check of our assumption, that the point reached in the limit $\lambda \rightarrow -\infty$ belongs to the curve $\sigma =0$. \\
%%%%%%%%%%  DOMINIO  %%%%%%%%%%%%
\begin{figure}[t]
\centering
%\vspace{1cm}
\includegraphics[width=.65\textwidth,angle=0]{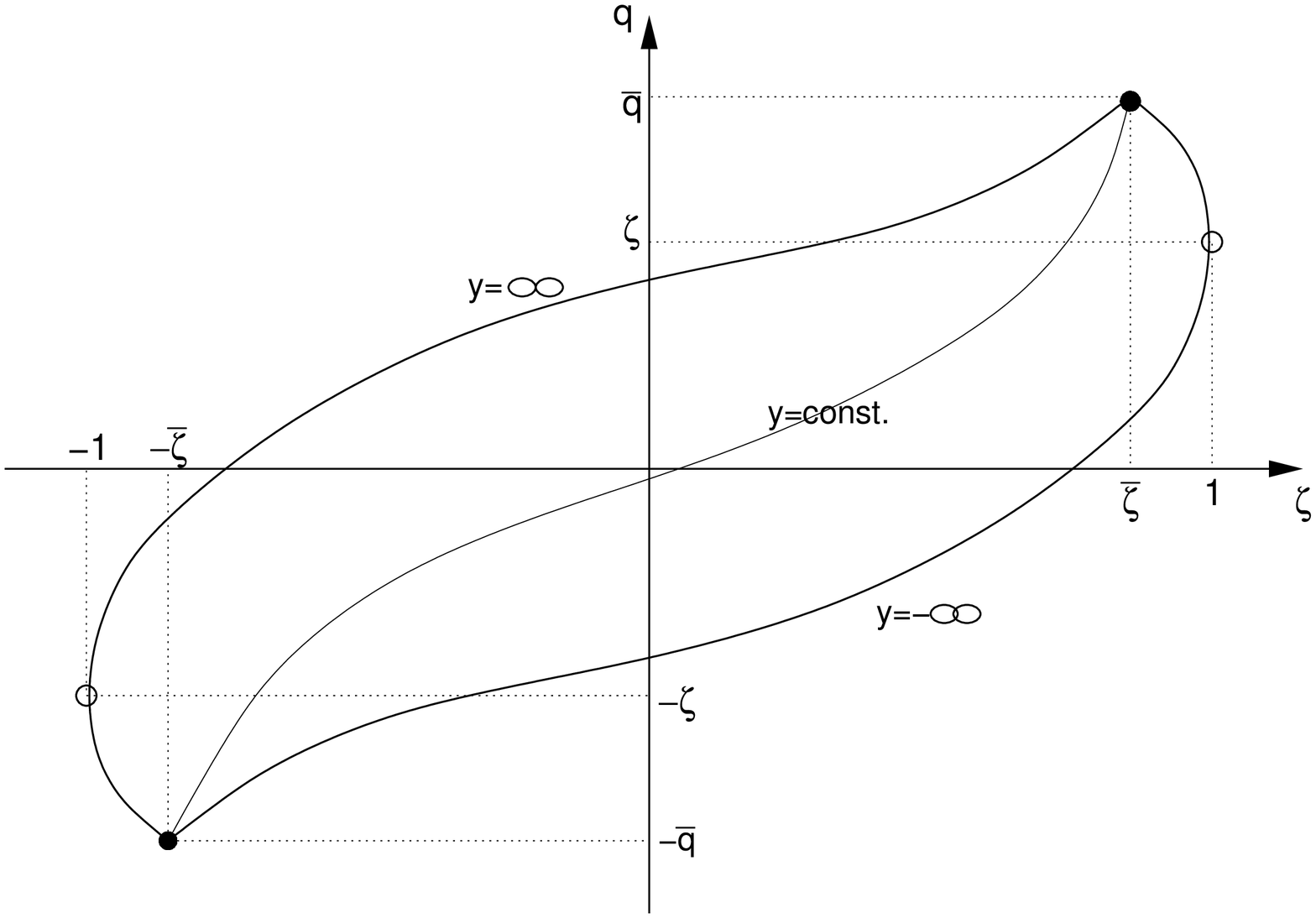}
\end{figure}
%%%%%%%%%%%%%%%%%%%%%%%%%%%%%%%%%%
A more accurate argument can be given in this way: one can look numerically at the curve $q(\zeta_s)$, parametrically in $\lambda$ at fixed $y$. The curve looks like the one reported in the figure above, and as $\lambda$ moves from $-\infty$ to $\infty$ the point in the $(\zeta_s,q)$ plane moves from $(\bar{\zeta}_s,\bar{q})$ to $(-\bar{\zeta}_s,-\bar{q})$ (black dots in the figure) as predicted by Eq.~(\ref{limit_lambda}). Now it is easy to show that
\beq
\begin{split}
\label{limit_y}
&\lim_{y \rightarrow \pm \infty} q(y,\lambda) = \mp \zeta \\
&\lim_{y \rightarrow \pm \infty} \zeta_s(y,\lambda) = \mp 1 \\
&\lim_{y \rightarrow \pm \infty} \sigma(y,\lambda) = 0 \\
\end{split}
\eeq
Then for $|y| \rightarrow \infty$ and $\lambda$ fixed the whole curves collapse on the white dots in the figure above. But it is also easy to show that for $y \rightarrow \infty$ and $\lambda=-y(1+\delta) \rightarrow -\infty$ the point goes on the upper border of the domain reported in the figure above, and moves from the left white dot to the upper black dot as $\delta$ moves from 0 to $\infty$ (the other branch, form the left white dot to the lower black dot, is obtained for $\delta$ going from $-2$ to $-\infty$). On the whole border of the domain we have $\sigma=0$ from Eq.~(\ref{limit_y}); then $(\bar{\zeta}_s,\bar{q})$ is exactly the point in which $q$ is maximum on the curve $\sigma=0$.

%%%%%%%%%%%%%%%%%%%%%%%%%%%%%%%%%%%%%%%%%%%%%%%%%%%%%%%%%%%%%%%%%%%%%%%%%%%
%                             REFERENCES
%%%%%%%%%%%%%%%%%%%%%%%%%%%%%%%%%%%%%%%%%%%%%%%%%%%%%%%%%%%%%%%%%%%%%%%%%%%

%%%%%%%%%%%%%%%%%%%%%%%%%%%%%%%%%%%%%%%%%%%%%%%%%%%%%%%%%%%%%%%%%%%%%%%%%%%%%%%%%%%%%%%%%%%%%%
%%%%%%%%%%%%%%%%%%%%%%%%%% FIGURES %%%%%%%%%%%%%%%%%%%%%%%%%%%%%%%%%%%%%%%%%%%%%%%%%%%%%%%%%%
%%%%%%%%%%%%%%%%%%%%%%%%%%%%%%%%%%%%%%%%%%%%%%%%%%%%%%%%%%%%%%%%%%%%%%%%%%%%%%%%%%%%%%%%%%


\begin{thebibliography}{99}

\bibitem{land_angell}
C.A.~Angell,
Science {\bf 267}, 1924 (1995).

\bibitem{land_sastry}
S.~Sastry, P.G.~Debenedetti and F.H.~Stillinger,
Nature {\bf 393}, 554 (1998).

\bibitem{land_sastry2}
S.~Sastry,
Nature {\bf 409}, 164  (2001).

\bibitem{land_keyes}
T.~Keyes and J.~Chowdhary,
Phys. Rev. E {\bf 65}, 041106 (2002).

\bibitem{deb_nature}
P.G.~Debenedetti and F.H.~Stillinger,
Nature {\bf 410}, 259 (2001).

\bibitem{land_buc}
S.~B\"uchner and A.~Heuer,
Phys. Rev. Lett. {\bf 84}, 2168 (2000).

\bibitem{sorin}
S.~Tanase-Nicola and J.~Kurchan, cond-mat/0302448.

\bibitem{stillinger}
F.H. Stillinger and T.A. Weber,  {Phys. Rev. A} {\bf 25}, 978 (1982);
Ibidem, {Science} {\bf 225}, 983 (1984); 
F. H. Stillinger, { Science}, {\bf 267}, 1935 (1995).

\bibitem{fs_entropy}
F.~Sciortino, W.~Kob, and P.~Tartaglia,
Phys. Rev. Lett. {\bf 83}, 3214 (1999).
E.~La Nave, S.~Mossa, and F.~Sciortino
Phys. Rev. Lett. {\bf 88}, 225701 (2002).

\bibitem{fabr}
G.~Fabricius and D.A.~Stariolo,
Phys. Rev. E {\bf 66}, 031501 (2002).

\bibitem{la_98}
L.~Angelani, G.~Parisi, G.~Ruocco, and G.~Viliani,
Phys. Rev. Lett. {\bf 81}, 4648 (1998).

\bibitem{donati}
C.~Donati, F.~Sciortino, and P.~Tartaglia,
Phys. Rev. Lett. {\bf 85}, 1464 (2000).

\bibitem{fs_aging}
F.~Sciortino and P.~Tartaglia,
Phys. Rev. Lett. {\bf 86}, 107 (2001).

\bibitem{FDTgen}
L.~Cugliandolo and J.~Kurchan, Phys.~Rev.~Lett. {\bf 71}, 173 (1993); 
Philosophical Magazine {\bf 71}, 501 (1995).
J.P.~Bouchaud, L.~Cugliandolo, J.~Kurchan, M.~M\'ezard, 
in ``Spin Glasses and Random Fields'', A.P.~Young~ed., World Scientific, 1997.

\bibitem{Cukupe}
L. F. Cugliandolo, J. Kurchan and L. Peliti,
Phys. Rev. E 55, 3898-3914 (1997).

\bibitem{cugliandolo2}
L.~Cugliandolo, ``Dynamics of glassy systems'', Lecture notes, Les
Houches, July 2002, cond-mat/0210312.

\bibitem{cugliandolo1}
J.P.~Bouchaud, L.~Cugliandolo, J.~Kurchan, M.~M\'ezard, Physica A {\bf 226}, 243 (1996);

\bibitem{Edwards}  
S. F. Edwards in ``Disorder in condensed matter physics'',
Oxford Science Publications, 1991,
and in ``Granular matter: an interdisciplinary approach'',
A. Mehta ed., Springer-Verlag, New York, 1994;
J. Kurchan, in ``Jamming and rheology: constrained dynamics in microscopic and 
macroscopic scales'', ITP, Santa Barbara, 1997, ed.~S.~F.~Edwards~{\it et al}, cond-mat/9812347. 

\bibitem{keyes_inm}
T.~Keyes,
J. Phys. Chem. {\bf 101}, 2921 (1997).

\bibitem{keyes_vari}
T.~Keyes, J.~Chem.~Phys.~{\bf 103}, 9810 (1995);
T.~Keyes, G.V.~Vijayadamodar, and U.~Zurcher, J.~Chem.~Phys.~{\bf 106}, 4651 (1997);
W.X.~Li and T.~Keyes, J.~Chem.~Phys.~{\bf 111}, 5503 (1999);
T.~Keyes, J.~Chowdhary, and J.~Kim, Phys.~Rev.~E~{\bf 66}, 051110 (2002).

\bibitem{noi_sad}
L. Angelani, R. Di Leonardo, G. Ruocco, A. Scala and F. Sciortino,
Phys. Rev. Lett. {\bf 85}, 5356 (2000);
L.~Angelani, R.~Di Leonardo, G.~Ruocco, F.~Sciortino, and A.~Scala, 
J. Chem. Phys. {\bf 116}, 10297 (2002);
L.~Angelani, G.~Ruocco, M.~Sampoli and F.~Sciortino,
submitted to J.~Chem.~Phys.

\bibitem{cav_sad}
K. Broderix, K.K. Bhattacharya, A. Cavagna, A. Zippelius and I. Giardina,
Phys. Rev. Lett. {\bf 85}, 5360 (2000).

\bibitem{sad_1}
J.~Chowdhary and T.~Keyes,
Phys. Rev. E {\bf 65}, 026125 (2002).

\bibitem{doye}
J.P.K. Doye and D.J. Wales,
J. Chem. Phys. {\bf 116}, 3777 (2002).

\bibitem{sad_3}
P.~Shah and C.~Chakravarty,
J.~Chem.~Phys.~{\bf 115}, 8784 (2001);
P.~Shah and C.~Chakravarty,
Phys.~Rev.~Lett.~{\bf 88}, 255501 (2002);
P.~Shah and C.~Chakravarty,
J.~Chem.~Phys.~{\bf 118}, 2342 (2003).

\bibitem{grig}
T.S.~Grigera, A.~Cavagna, I.~Giardina, and G.~Parisi,
Phys. Rev. Lett. {\bf 88}, 055502 (2002).

\bibitem{sadBLJ}
M.~Sampoli, P.~Benassi, R.~Eramo, L.~Angelani, G.~Ruocco,
J.~Phys.:~Condens.Matter~{\bf 15}, S1227 (2003).

\bibitem{parisi_boson} G.~Parisi, cond-mat/0301284, to be published by 
Journal of Physics; cond-mat/0301282, Physica A (in press).

\bibitem{mct}
W. G\"otze,
J. Phys.: Condens. Matter {\bf 11}, A1 (1999).

\bibitem{pettini&co}
R.~Franzosi, M.~Pettini, and L.~Spinelli,
Phys.~Rev.~Lett. {\bf 84}, 2774 (2000);
L.~Casetti, M.~Pettini, and E.G.D.~Cohen, 
Physics Reports {\bf 337}, 237 (2000);
L.~Casetti, E.G.D.~Cohen, and M.~Pettini, 
Phys.~Rev.~E {\bf 65}, 036112 (2002).

\bibitem{noieulero}
L.~Angelani, L.~Casetti, M.~Pettini, G.~Ruocco, F.~Zamponi, cond-mat/0205483, to be published on Europhys.~Lett.

\bibitem{Kula}
J. Kurchan and L. Laloux, J. Phys. A: Math. Gen. {\bf 29}, 1929 (1996).

\bibitem{Bimo}
G. Biroli and R. Monasson, Europhys. Lett. {\bf 50}, 155 (2000).

\bibitem{cav2}
A.~Cavagna, I.~Giardina, and G.~Parisi, Phys.~Rev.~B {\bf 57}, 11251 (1998);
A.~Cavagna, I.~Giardina, and G.~Parisi, J. Phys. A: Math. Gen. {\bf 34}, 5317 (2001).

\bibitem{nota1}
Obviously, if $\psi = 2\pi n/k \neq 0$, the unbroken simmetry is $\varphi \rightarrow 2\psi - \varphi$.

\bibitem{ZJ}
J.~Zinn-Justin, ``Quantum Field Theory and Critical Phenomena'', Clarendon Press, Oxford, 1989.

\bibitem{Risken}
H.~Risken, ``The Fokker-Planck Equation'', Springer-Verlag, Berlin, 1984

\bibitem{nota2}
While for $\zeta = 0$ Eq.~(\ref{effdynresc}) makes no sense ($\tilde{T}$ is infinite), we obtain the correct result for the diffusion constant in the paramagnetic phase substituting $\zeta = 0$ in the expression obtained in the magnetic phase. This will not be the case for the relaxation times of the correlation functions.

\bibitem{nota_sim}
The algorithm used is a "Prime Modulus M Multiplicative Linear Congruential 
Generator" a modified version of the random number generator by Park and Miller 
in "Random Number Generators: Good Ones Are Hard to Find", CACM, October 1988, 
Vol. 31, No. 10.

\bibitem{parisi_boson2} Another denomination is ``generalized inherent structures'', see Ref.~\cite{parisi_boson}.

\bibitem{noidinamica}
L.~Angelani, G.~Ruocco, F.~Zamponi, cond-mat/0212098, to be published in J.~Chem.~Phys.

\bibitem{chi_e_SUSY}
J.~Kurchan, J.~Phys.~A:~Math.~Gen.~{\bf 24}, 4969 (1991);
A.~Cavagna, I.~Giardina, G.~Parisi, M.~M\'ezard, J.~Phys.~A:~Math.~Gen.~{\bf 36}, 1175 (2003)
and references therein.

\end{thebibliography}
\end{document}